\documentclass[11pt, a4paper, twoside]{article}

\usepackage{graphicx}
\usepackage[usenames]{color}

\usepackage[english]{babel}

\usepackage{amssymb,latexsym,amsmath}

\usepackage[numbers]{natbib}

\newcommand{\const}{\textit{Const}}
\newcommand{\dd}{\mathrm{d}}
\newcommand{\gdg}{\texttt{GADGET}}
\newcommand{\tree}{TREE}
\newcommand{\fof}{\texttt{FOF}}

\newcommand{\vs}{\emph{vs.}}
\newcommand{\etc}{\emph{etc}}

\newcommand{\Vector}[1]{{\bf #1}}
\newcommand{\Msun}{\,\mathfrak{M}_\odot}

\newcommand{\HII}{H$\,$II}

\newcommand{\HH}{H$_2$}

\newcommand{\eref}[1]{\mbox{(\ref{#1})}}
\newcommand{\scient}[2]{#1\cdot\!10^{#2}\,}
\newcommand{\myfootnote}[1]{\!\!\protect\footnote{#1}}

\newcommand{\mult}{\!\times\!}
\newcommand{\simi}{\sim\!}
\newcommand{\isimi}{\!\sim\!}
\newcommand{\myindex}[1]{${}^{#1}$}

\title{{\bf The study of colliding molecular clumps evolution}}

\author{\myindex{1}Vinogradov S.B., \myindex{2,1}Berczik P.P. \bigskip \\
        \small \myindex{1}Main Astronomical Observatory \\
        \small Ukrainian National Academy of Sciences \\
	\small 27 Zabolotnoho str., Kiev, Ukraine, 03680 \\
	\small e-mail: \texttt{vin@mao.kiev.ua} \bigskip \\
        \small \myindex{2}Astronomisches Rechen-Institut (ARI) \\
        \small Zentrum fur Astronomie Univ. Heidelberg (ZAH) \\
	\small Monchhofstrasse 12-14, 69120 Heidelberg, Germany \\
        \small e-mail: {\tt berczik@ari.uni-heidelberg.de}
	}
 
\small \date{15 August 2006}

\begin{document}

\pagenumbering{roman}

\maketitle

\begin{abstract}

\textit{The results of study of the gravitational fragmentation in the interstellar 
medium (ISM) by clump-clump collisions are presented. We suggest, that 
collision of clumps, that are subparts of Giant Molecular Clouds (GMC) may 
be on of the basic mechanism, which result to ISM fragmentation and define 
the dynamical as well as statistical characteristics (e.g. the mass spectra) 
of protostellar condensation.}

\textit{In the present paper, we describe our $3D$ SPH-modeling, in isothermal 
approximation, of supersonic collisions of two identical clumps 
with a few variants of initial impact parameters ($\beta$), that cover the wide 
range.}

\textit{Our results shown, that at all $\beta$ in system began intensive 
fragmentation. The resulting fragments mass function depend from initial 
impact parameter. The obtained mass spectra have the slopes in a good enough 
agreement with observational data for our Galaxy -- especially for large 
impact parameters, which are more realistic as for large clumps ensembles.}

\vspace{5mm}

\textsl{KEY WORDS} \hspace{3mm} ISM, fragmentation, molecular clouds collisions, 
clumps mass spectra, SPH method

\end{abstract}

\pagenumbering{arabic}

\section{Introduction} \label{S:Intro}

Nearly the half mass of gas in our Galaxy are concentrated in giants molecular 
clouds (GMC) with typical sizes in tens parsecs, masses $10^5-10^6 \Msun$ and 
temperatures $T \lesssim 20 \textrm{K}$ \citep{Bochkarev81, Shu87,Ferriere}.
The mean number density in GMC is equal to $10^2-10^3 \; \textrm{cm$^{-3}$}$, 
but in some regions it can reaches up to $10^6 \; \textrm{cm$^{-3}$}$ \citep{Shu87}. 
There are about 4000 such objects in Galaxy and this objects mainly concentrate 
in the Galaxy disc \citep{Bochkarev81, Shu87, Ferriere}.

The GMC tightly connect with some other objects in our Galaxy -- such as 
IR-sources, masers, compact \HII{} zones, T Tau stars, Herbig-Haro objects, 
O-association \etc. This connection show the clear evidence of star formation 
process in GMC at present.

The hight resolution radio observations of GMC in molecular lines, show their clear 
fragmentary structure on all hierarchy levels down to parsec's fractions. In first
approximation we can describe the GMCs as a complex of smaller and denser 
condensations (clumps), embedded into tenuous gas medium.

The typical clump have a mass in $10^{3} - 10^{4} \; \Msun$ and size in a few 
parsecs. It content yet more small and dense cores with masses in several solar 
mass and sizes in parsec's fraction \citep{Shu87}. As a good examples of such 
clouds complexes we can brought the $\rho$ Oph \citep{Motte1998}, M17 SW 
\citep{Stutzki_Gusten1990} and Tauri molecular clouds complex \citep{AK90}.

The numerous observational data about molecular clouds indicate the complex 
gas flows and clumps chaotic motions inside GMC. There are strong turbulence 
inside clumps, but tenuous gas "background" have a more regular kinematics. 
The dependence of velocity dispersion $\sigma$ from size of consider region 
can be approximated by the power law $\sigma \propto L^{0.38}$ \citep{Larson81}. 
Here $\sigma$ is total 3D-rms velocity deviation for all internal motions 
including large-, small-scale and thermal gas motions (for \HH{} clumps at 
temperature in a few tens of Kelvins the mean quadratic velocity of thermal 
motions is negligible by comparison with two first components). $\sigma$ is 
about $1 \;\textrm{km/s}$ for $R \approx 2 \;\textrm{pc}$ or $\mathfrak{M} 
\approx 2000 \;\Msun$. 

This discovered dependence is very similar to Kolmogorov law for distribution 
of velocity dispersion at subsonic turbulence ($\sigma \propto L^{0.33}$) and 
it is possibly, that all seen us interstellar movements is parts of a single 
hierarchy of interstellar turbulence.

At that time the rms velocity dispersion of molecular clouds and clumps itself 
have mainly flat spectrum in wide mass range (up to four orders) with average 
value around $5 - 10 \;\textrm{km/s}$ \citep{Casoli_etal84, Stark84}.

Usually, the GMC, having masses $10^5\!-\!10^6 \Msun$, may contain a tens of 
clumps with typical masses by order $10^3\!-\!10^4 \Msun$ \citep{Elmegreen85}. 
Of course, chaotically moving at supersonic velocities, they can collide 
each with other. Many authors consider this mechanism as one of the possible 
source of ISM fragmentation and formation of initial mass function (MF) of 
protostellar condensations. 

It is interesting to note, that as GMC, as clumps and cores inside its have 
almost the same mass spectra in mass range from $10^{-4}$ to $10^{4} \; \Msun$ 
\citep{Kramer_etal1998}.  This may point to common mechanism of their formation. 
For example, just such result was obtained in special N-body 
modeling of collisional buildup of molecular clouds \citep{Das_and_Jog96}, 
that gave observational mass- and velocity spectra for obtained objects.

So, interclouds collisions have a great interest for its study, including a 
numerical simulations. First, two-clouds isothermal collisions are studied: 
for identical rotating clouds \citep{LH88}, non-rotating clouds with different 
masses \citep{LMPS85} and collisions of identical clouds with taken into account 
cooling/heating processes \citep{MV88, ML2000}. 

Simulations with using great particle numbers ($\simi 10^5-10^6$) and, 
therefore, with hight resolution require sufficient CPU consumptions even for 
supercomputers. In consequence of that, they carrying out either without 
integrating of energy equation, but with barotropic state equation  
$p= p(\rho)$ \citep{Bhattal98}, or simply in isothermal approximation 
\citep{Gittins03}.

In \citet{MV88} and \citet{Bhattal98} the system of 48 and 1000 clouds 
correspondingly was considered. But even in such rich complexes the main 
role play just a pair collisions \citep{Gittins03}.

The most of such works were carried out without quantitative analysis of 
fragmentation. Thus, only in \citet{Bhattal98} and \citet{Gittins03} the masses 
of fragments are evaluated. Additionally, in the last work qualitative analysis 
of IMF is given for which the fragments are defined as a single sink particles 
without any internal structure, \etc{}  \citep[see ][ for more details]{Bate95}.

In our work we carried out simulation of isothermal supersonic collisions of 
two identical molecular clumps with aim of more detailed fragmentation study 
and mass distribution analysis of obtained fragments. For fragments finding 
we apply the modified variant of well-known "friends-of-friends" algorithm (FOF).

\section{The Method} \label{S:Method}

\subsection{Hydrodynamics} \label{SS:Hydrodynamics}

For gas process modeling we use the \emph{Smoothed Particle Hydrodynamics}
method (SPH), which was independently suggested by \citet{L77} and \citet{GM77}.

As in the well-known N-body algorithm, modeling system in SPH is represented 
by a set of computational particles which characterized by mass, position, 
linear and angular momentums and energy. Besides, each particle is diffusive 
one and additionally have some hydro- and thermodynamical parameters (such as 
density and thermal energy), defined for its "center" by interpolating from 
nearest neighborhood. Partially overlapping, SPH-particles gives the average 
weighing contributes to local properties of medium, according with their 
internal distribution profile. 

So, the smoothing estimate of any physical parameter $f(\Vector{r})$ in SPH 
is:
\begin{equation} \label{E:SPH-Int}
\langle f(\Vector{r}) \rangle= \int f(\Vector{r}') W(\Vector{r}-\Vector{r}',h)\;d\Vector{r}'
\approx \sum_{j=1}^{N} m_j \frac{f(\Vector{r}_j)}{\rho_j} W(\Vector{r}_i-\Vector{r}_j,h_i),
\end{equation}
\noindent where the integral is taken over entire space and summation is over 
all $N$ particles, $h$-- so-called smoothing length, $W(\Vector{r}-\Vector{r}',h)$
-- interpolating smoothing kernel, which must satisfy to two simple conditions:
$
\int W(\Vector{r}-\Vector{r}',h)\;d\Vector{r}' \equiv 1 \; \textrm{and} \; 
\lim_{h \to 0}W(\Vector{r}-\Vector{r}',h)\!=\! \delta(\Vector{r}-\Vector{r}').
$

Yet another SPH features is that it need not in finite-differences for gradients
calculating. Instead of this we must only differentiate the smoothing kernel 
and then smooth interesting parameter as in \eref{E:SPH-Int}:
\begin{equation} \label{E:SPH-Grad-Int}
\langle \nabla f(\Vector{r}) \rangle= \int f(\Vector{r}') \nabla W(\Vector{r}-\Vector{r}',h)\;
                                      d\Vector{r}'
\approx 
\sum_{j=1}^{N} m_j \frac{f(\Vector{r}_j)}{\rho_j} \nabla W(\Vector{r}_i-\Vector{r}_j,h_i).
\end{equation}

The concrete form of kernel is chosen from reason, that non-zero contribution to 
local hydrodynamical properties must take only the nearest particles. Smoothing 
length $h$ define a scale of smoothing and so the spatial resolution.

The great flexibility of SPH is provided by absence of any restriction 
neither on system geometry, nor on admissible range of varying of dynamics 
parameters, easiness of kernel change (that is equal to change of finite-difference 
scheme in grid methods), variety of equation symmetrization means, artificial 
viscosity forms, \etc.

Being hydrodynamics extension of earlier N-body method, SPH can naturally include 
it, as a one of its parts, for modeling collisionless components (as a dark matter).
For handle with selfgravity field SPH use either grid methods to solve of Poisson's 
equation, or \tree-algorithm \citep{HK89}, or, at last, direct summation of all
particle-particle interactions by special hardware GRAPE. Also, SPH may be easy 
combined with other optional algorithms such as starformation, feedback \etc{} 
\citep[see for example ][]{Berczik99, Berczik2000, Berczik2003, Spurzem2004, 
CLC96}.

For more details see the good reviews by \citet{M92}, \citet{BK93}, \citet{HK89}, 
\citet{HV91}, and other works such as \citet{Lomb99}, \citet{Thaker}, 
\citet{Berczik2000}, \citet{BK98}, \citet{CLC96}, \citet{ML85}, \citet{SM93}  
and of course \citet{Spring}.

We used the parallel version of freely available SPH code \gdg{} version 1.1
\myfootnote{\texttt{http://www.mpa-garching.mpg.de/gadget}}
({\bf{GA}}\textit{\-la\-xi\-es wi\-th} {\bf{D}}\textit{ark mat\-ter and } 
{\bf{G}}\textit{as int}{\bf{E}}\textit{r\-ac}{\bf{T}}) \citep{Spring}, that 
destinate for numerical SPH/N-body simulating both of isolated self-gravitating 
system and cosmological modeling in comoving coordinates (with/with\-out periodical
boundary conditions). The \gdg{} v 1.1 distributive contain two codes: serial one, 
for workstations, and parallel -- for supercomputers with distributed memory or 
PC-clusters with shared memory.  Code is written on ANSI C and use standard  MPI 
library (\emph{Message Passing Interface}) 
\myfootnote{\texttt{http://www-unix.mcs.anl.gov/mpi/}}
for parallelization.

\gdg{} allow up to four types of N-body particles for collisionless fluids
and one type of SPH particles for gas in modeled system. The main code features
are spline-based smoothing kernel \citep{ML85}, which is second order accuracy
(in since that $<\!\!f(\Vector{r})\!\!> = f(\Vector{r}) + O(h^2)$), shear-reduced 
form of generally accepted Monaghan artificial viscosity \citep{Lomb99, Thaker, 
M92}, Burnes \& Hut (BH) \tree-algorithm for gravity calculation, leapfrog scheme 
as a time integrator \citep{HV91, BK93} and so. 

The code is completely adaptive in time as well as in space, owing to algorithms 
of individual time-steps and smoothing length for each particle. The last feature 
signify, that the number of nearest smoothing neighbors, which must influence on 
local medium properties is just constant in serial code version and nearly constant 
(in small enough bands in a few percents) in parallel version \citep{Spring}.

The time-step for each active particle calculated as 
\begin{equation} \label{E:TS1} 
\Delta t_1= \frac{\alpha_{toll}}{|\Vector{a}_i|}.
\end{equation}
\noindent Additionally, there is yet another hydrodynamical criterion specially
for SPH particles:
\begin{equation} \label{E:TS2} 
\Delta t_2= \frac{\alpha_C h_i}{h_i\, |(\nabla \Vector{v})_i| + 
            \max (c_i,\, |\Vector{v}_i|)\cdot (1+0.6\,\alpha_{visc})},
\end{equation}
\noindent and taken the minimal from both.
(here \Vector{a} is acceleration, $\alpha_{toll}$ -- dimensional accuracy factor 
\citep[see][]{Spring}, $\alpha_C$ -- analog of Courant factor, $\alpha_{visc}$ -- 
artificial viscosity factor, $c$ -- sound speed). Besides, time-steps are bounded
by their lower and upper limits: $\Delta t_{min}$ and $\Delta t_{max}$.

For gravity calculation in \gdg{} were employed the standard BH oct-\tree, which 
is built for each particles type separately. For SPH particles it used also for 
neighbors search. Gravity potential expanded up to quadrupole order. As a 
cell-opening criterion may be used either standard BH-criterion \citep{BH86}, or 
original one, based on approximative estimate of force error, caused by multiple 
expansion oneself \citep{Spring}.

\subsection{Fragments finding} \label{SS:FOF}
As our purpose was to study the fragmentation in colliding clumps, we has not 
only to simulate the collision, but also to find an appropriate method for the
determination of the extent of medium fragmentation and for the statistical 
analysis of fragments properties.

We decided in favor of the well-accepted "friends-of-friends" algorithm 
(hereafter \fof). It was initially proposed for the separation of galactic groups
and clusters in the spectroscopic redshifts surveys \citep{HG82}. This simple and 
versatile algorithm is still widely used \citep{Frederic94, Botzler2003}. As 
original \fof{} method handled with points on coelosphere and distances between 
them it can be easily adapted for analyzing the results of numerical N-body/SPH
experiments \citep{DEFW85, BE87}.

For each particle, that has not yet been subjected to the analysis procedure, 
all its neighbors located inside the sphere of a certain so-called "search-linkage"
radius $\delta$ are found. When there are no such neighbors, the particle is 
write off as isolated one (or as a background particle), otherwise this procedure
is recursively repeat to all its neighbors are found, neighbors of neighbors and 
so on, until the whole list is exhausted. Then all the particles thus found are
included in a cluster (fragment) if their amount is not below some threshold 
$N_{min}$. In other words, if any two particles are separated by less then $\delta$,
they both belong to the same group (but only in the case when the whole group 
will collected no less then $N_{min}$ members, else all these particles are 
reckoned as belonging to the background).

The advantages of the \fof{} algorithm are its coordinate-free nature and so 
absence of any meshes and independence from problem geometry; no \textit{a priori}
assumptions are made as to properties of the sought-for groups of particles
(such as shape, symmetry, density profile \etc.), which is very convenient in
the work with 3D models.

The algorithm has only two "adjustment parameters": $\delta$ and $N_{min}$.
The former parameter control the compactness and extension of the sought-for
fragments, while the later one rejects random small groups of closely located
particles like pairs and triples (the statistics suggests, that they comprise
about $80-85$ \% of all groups found). Generally spiking, $N_{min}$ should be 
as small as possible, but not smaller then the number $N_B$ of smoothing 
neighbors.

Choosing of "search-linkage radius" (SLR) $\delta$ is not all the clear, but 
obviously it should not be too large. It often taken in terms of the mean 
interparticles distances (if that's case the geometric mean is better, than
arithmetic mean, because the existence of very distant, very compact groups
can result in unjustifiably large $\delta$). In this case, however, at different 
moments the fragments are separated with the use of different scales, i.e., 
$\delta$ will depend from time. In addition, any mutual fragments moving off 
will influence the results of analysis. So, it is necessary to use the most 
stable criterion (as regards to interparticles distances varying) for SLR 
choosing, or set it by hand.

Apart from all, the simple \fof{} method have some drawbacks. For example two 
groups which linked by a small thread will be identified as a single structure;
or when there are several groups with strong different compactness and/or with 
internal fragmentation, the algorithm may either do not recognize the internal 
structure of some fragments, or it can let go some less compact groups at smaller 
$\delta$.

We modified the standard \fof{} algorithm to avoid such situations, by manual
setting up of a few fixed SLR ${\delta_i}$ and multistep fragments searching. 
All $\delta_i$ are consequently applied by they increasing to nonidentified 
particles (i.e. writing off from previous step). So, first the most compact 
fragments will be detected with the smallest $\delta$; next the less compact
fragments are found with the new SLR among remained particles and so on.

We took four radii $\delta= 0.001, 0.005, 0.01, 0.05\; \textrm{pc}$. The largest  
radius is approximately equal to the gravitational smoothing length (see below),
and smallest radius was chosen for the densest fragments recognizing as a separate
ones. The threshold particles number $N_{min}$ was equal to number of smoothing 
neighbors $N_B= 40$.

\section{Scaling and initial model} \label{S:InitConf}

\gdg{} allow to use arbitrary user's system of units, including dimensionless 
variables. As a base units, are units of mass, length and velocity. We choose
$$
[\mathfrak{M}]= 1\Msun,\qquad [l]= 1\:\textrm{pc},\qquad [v]= 1\:\textrm{km/s}.
$$
Following this, unit of time is defined as $[t] \!= [l][v]^{-1} \!\approx 0.978 
\;\textrm{Myr}$, density -- $[\rho]= [\mathfrak{M}][l]^{-3} \approx 
\scient{6.77}{-20} \; \textrm{kg/cm$^{3}$}$ (that correspond to number density 
$17.6 \;\textrm{cm$^{-3}$} $, at choosen mean molecular weight -- see below). 
The temperature measured in Kelvins, but everywhere 
in code use the internal specific energy $u$ instead of temperature, which is 
measured in $[v]^2$. In chosen units gravitation constant is $G = \scient{4.3}{-3} 
\;\textrm{$\Msun^{-1}$ $\cdot$ pc $\cdot$ (km/s)}$, gas constant -- $\mathfrak{R}
= \scient{8.314}{-3} \textrm{$\Msun\cdot$ (km/s)$^{2} \cdot$ mol$^{-1} \cdot$ 
K$^{-1}$}$.

We consider four models of supersonic collisions of two identical molecular 
clumps with various impact parameters $\beta \equiv b/R= 0, \; 0.2, \; 0.5, \; 
0.75$ (were $b$ is linear shift of clumps centers, $R$ - their radius) and 
therefore with different initial angular moment.

\begin{figure}
\centering
\includegraphics[width=4.8cm]{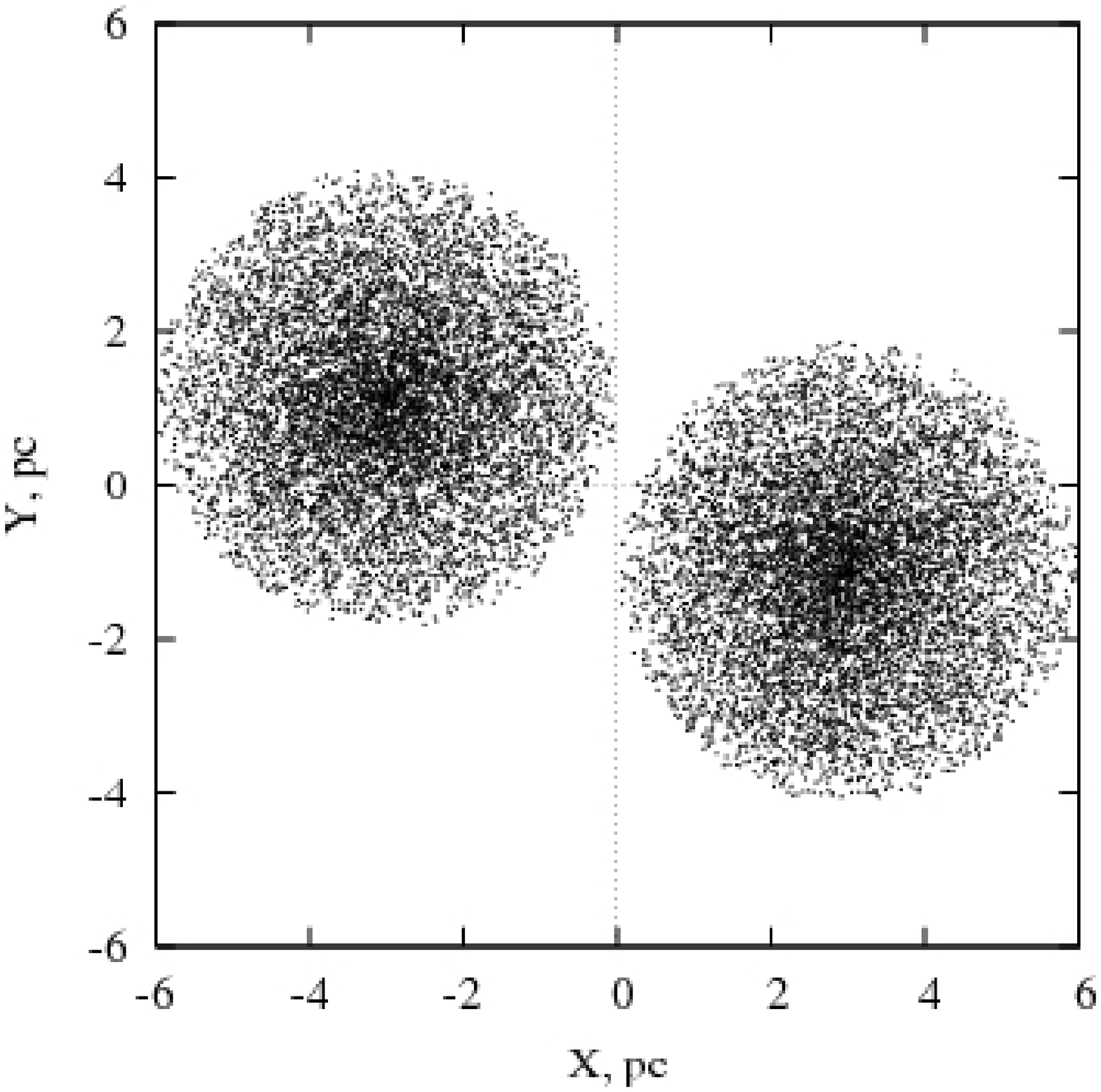} \hfill 
\includegraphics[width=4.8cm]{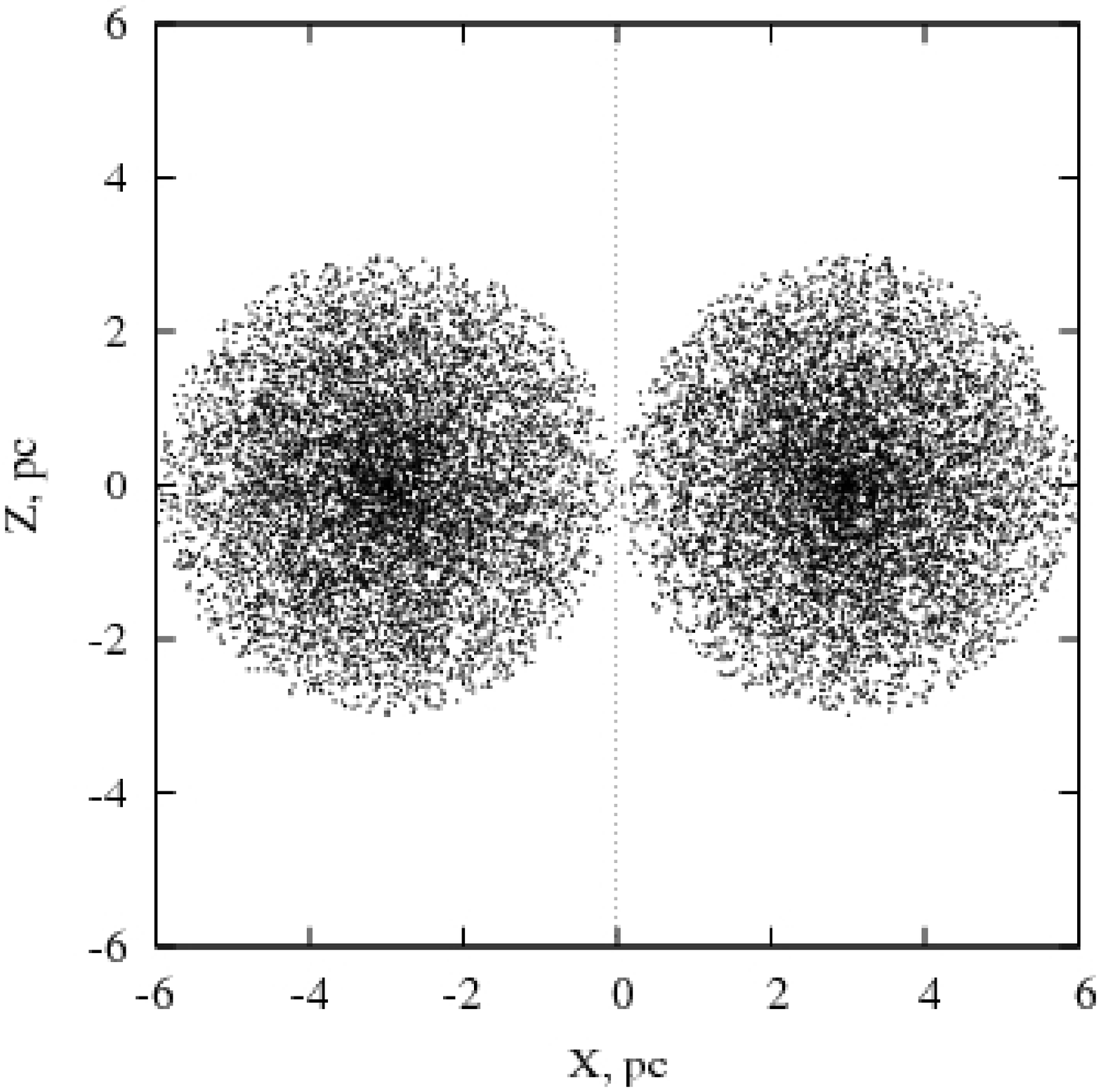} \hfill
\includegraphics[width=4.8cm]{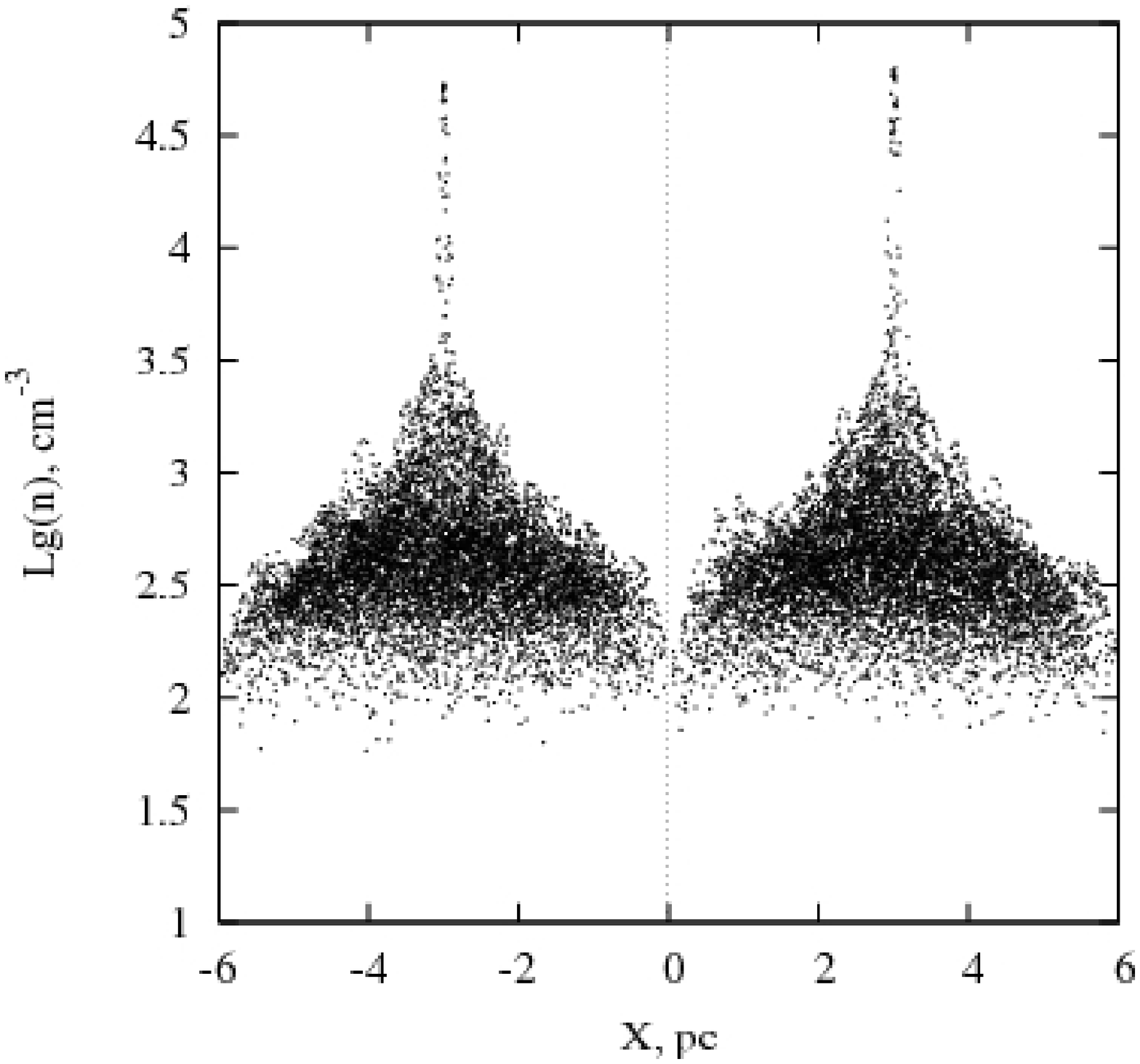}
\caption{Initial clouds position at $\beta= 0.75$.} 
\label{P:InitExample}
\end{figure}

Firstly, clumps' centers positioned at $\{-R; \frac{b}{2}; 0\}$ and $\{R; 
-\frac{b}{2}; 0\}$. There are no any internal gas motions in both clumps
initially, but they moves to each other along $X$-axis each with 
velocity $v$ (see Fig. \ref{P:InitExample} for example).

The mass $\mathfrak{M}$ in both clumps is distributed inside cut-off radius $R$ 
according to density profile
\begin{equation} \label{E:InitRho}
\rho= \frac{\mathfrak{M}}{2 \pi R^2}\: \frac{1}{r}.
\end{equation}
We choose the follow basic parameters for each clump: 
$$
\mathfrak{M}= 2000 \Msun, \quad 
R= 3 \:\textrm{pc}, \quad 
T= 20 \:\textrm{K}, \quad  
v= 5 \:\textrm{km/s}, \quad 
\mu= 2.3
$$
That is in good agreement with observational data about ISM in our Galaxy 
\citep{Shu87, WBM99, BW99, Ferriere}.

Additionally, each clump have further derived characteristics:
\begin{list}{$\blacktriangleright$}{\setlength{\itemsep}{0pt} 
                                    \setlength{\parsep}{0pt} 
			            \setlength{\itemsep}{1mm}
			            \setlength{\parskip}{0pt}} 
\item free-fall time: \dotfill $\tau_{f\!f} \approx 1.92 \:\textrm{Myr}$,
\item crossing time: \dotfill $\tau_{cross}= R/v \approx 0.59 \:\textrm{Myr}$,
\item isothermal sound speed: \dotfill $c \approx 0.27 \:\textrm{km/s}$,
\item mean number density (at given $\mu=2.3$): \dotfill $\bar{n} \approx 300 
      \:\textrm{cm$^{-3}$}$,
\item ratio of thermal energy to gravitational one: \dotfill $E_T/|E_G| \approx 
      0.09$,
\item Jeans' mass: \dotfill $\mathfrak{M}_J \approx \! 1\:\Msun$ in center and 
                            $\approx \! 10\:\Msun$ at edge.
\end{list}

For generation of initial configuration we divide up each clump into $100$ shells
of equal mass with boundary radii obeying to density profile \eref{E:InitRho}.
Next, we homogeneous and randomly distribute all particles in total number $N$ 
among this shells in corresponding shares independently for each clump. It is 
need to note, that then any ideal symmetry was wittingly excluded and there was 
slight departure in positions of clumps centers from theoretical ones.

The two set of calculations were carried out: one with total number of 
particles $N= 2 \mult 4000$ and second -- with $N= 2 \mult 8000$. In all 
calculations we used the fixed number of smoothing neighbors $N_B= 40$, 
$\alpha_{visc}= 0.75$, $\alpha_{toll}= 1.00 \:\textrm{km/s}$, $\alpha_C= 0.01$, 
$\Delta t_{min}= 10^{-6} \:\textrm{[t]}$, $\Delta t_{max}= 0.75 \:\textrm{[t]}$ 
and gravitation softening $\varepsilon= 0.03 \:\textrm{пк}$ 
\citep[see][]{Spring}.

For calculation of all models we had used the supercomputing facilities of
Hungarian National Supercomputing Center
\myfootnote{\texttt{http://www.iif.hu/szuper}} 
(SUN E10k/E15k supercomputers). 

\section{The Test Runs} 
\label{S:Tests}

Besides main simulations, we first fulfilled a few simple test-runs for 
study of code characteristics oneself. 

First test was adiabatic collapse of cold cloud with density profile 
\eref{E:InitRho} and homogeneous specific internal energy distribution 
$$
u = 0.05 \frac{G\mathfrak{M}}{R}
$$
in natural dimensionless units system  $G \!=\! \mathfrak{M} \!=\! R \!=\! 1$.

This test, named as "Evrard-test", was suggested in \citet{Evrard} and now became 
a standard tool for SPH-codes verification \citep{HK89, Thaker, SM93, CLC96, 
Spring}. It was calculated out with different particles number $N=(1, 2, 4, 8, 16, 
32) \!\cdot\! 10^3$ till final moment nearly equal to $3 \:\tau_{f\!f}$.

In the full agreement with cited works, at $t\approx0.7-1 \;\tau_{f\!f}$ central 
bounce was occurred and during outwards shock wave going, significant dissipation 
of kinetic energy into heat taken place. Nearly at $2.5 \;\tau_{f\!f}$ in whole 
system the virial equilibrium between thermal and potential energies $E_T \approx 
-E_G/2$ had been set. Also, test showed the excellent energy conservation 
$|\delta E/E| < 0.1 \%$ even at small $N$.

Generally speaking, the total CPU-consumption must determine by selfgravity 
calculation. Our results showed the typical for \tree-algorithms asymptotic of 
time consumption $\simi O(N\,log(N))$, but the hydrodynamical consumption is 
even greater then for gravity.

As the next test, we calculated out the spherical isothermal collapse of analogous 
cloud with mass $\mathfrak{M}= 2000 \Msun$, radius $R= 3 \:\textrm{pc}$, but with 
temperature $T= 20 \:\textrm{K}$, as in main simulations. Particles number was set 
to $N=4000$. After $1.56 \,\textrm{Myr}$, that correspond to $0.81\, \tau_{f\!f}$, 
central dense core with size of $\simi 0.03 \:\textrm{pc}$ was formed close to 
center 
\myfootnote{due to character of initial configuration building, there is no exact 
coincidence with geometrical center (see sec. \ref{S:InitConf} for more details.)}
with thin extensive shell around, that have self-similar density distribution 
$\rho \propto r^{-2}$ \citep[see][]{Larson69}.

The third test was fulfilled for check-up of comparative role of each time-step 
criterion \eref{E:TS1} and \eref{E:TS2} and for choice of available accuracy 
factors. We calculate adiabatic collapse from "Evrard-test" with $32000$ particles
once with fixed $\alpha_{toll}= 0.05 (G\mathfrak{M}/R)^{1/2}$ and various $\alpha_C=
0.1, \; 0.05, \; 0.01, \; 0.001$, and another time -- the wrong way, with fixed 
$\alpha_{C}= 0.01$ and various $\alpha_{toll}= 0.02, \; 0.05, \; 0.07$. With 
changing of Courant factor $\alpha_C$ the CPU-consumptions was increasing pro rata 
to its decreasing, that point to determinant role of criterion \eref{E:TS2}. Since 
energy error with $\alpha_{C}$ decreasing approach to some limit ($\simi 0.05 \%$), 
we choose -- as a available from ratio accuracy/consumption -- its value $\alpha_C= 
0.01$. At such choice, variation of $\alpha_{toll}$ even does not any affect neither 
on accuracy, nor on time-consumption. 

And in the end, we calculate one of clumps collisions -- with $N= 2 \mult 4000$, 
$v= 5 \:\textrm{km/s}$, $\beta= 0.5$ during $1.3 \;\textrm{Myr}$ on 1, 2, 4, 8, 16 
and 32 SUN UltraSPARC III (1050 MHz) processors for verify of code parallelization.
\myfootnote{for single CPU calculation the serial \gdg{} version was used.}
(see tab. \ref{T:CPUdata1}) 

In general case, any parallel code may be split to so-called consecutive and 
parallel parts. Therefore it is easy to characterize a quality of code 
parallelization by relative share of this part in total time-consumption. For 
example, on parallel part fall
\[
\tau_p = \frac{1-1/k}{1-1/N_p},
\]
\noindent where $N_p$ -- processors number, $k>1$ -- ratio of durations of task
work on single CPU and on $N_p$ processors. For \gdg{} this value is greater then 
93\%. Also, we can evaluate the effectiveness of CPU loading, as $\epsilon = 
k/N_p$ (tab. \ref{T:CPUdata1}).

\section{The Results}
\label{S:Results}

\underline{\textbf{\textsl{Head-on collision ($\beta=0$)}}} \hspace{7mm}
Because of extensively CPU-consumption this model was traced till $2.885 
\;\textrm{Myr}$ only. However, already at $1 \;\textrm{Myr}$ for $N\!=\!2\mult8000$ 
and at $1.76 \;\textrm{Myr}$ for $N\!=\!2\mult4000$ the number density of central 
object amounted of $\scient{4.5}{10} \;\textrm{cm$^{-3}$}$ and exceeded the 
"opacity limit" ($10^{-13} \;\textrm{g cm$^{-3}$} \cong \scient{2.6}{10} 
\;\textrm{cm$^{-3}$}$, see \citet{MI2000}). So, there was no reason to continue 
simulation in isothermal approximation at last for this object.

\begin{figure}[Htbp]
\vspace{-2cm}
\centering
\includegraphics[width=5.0cm]{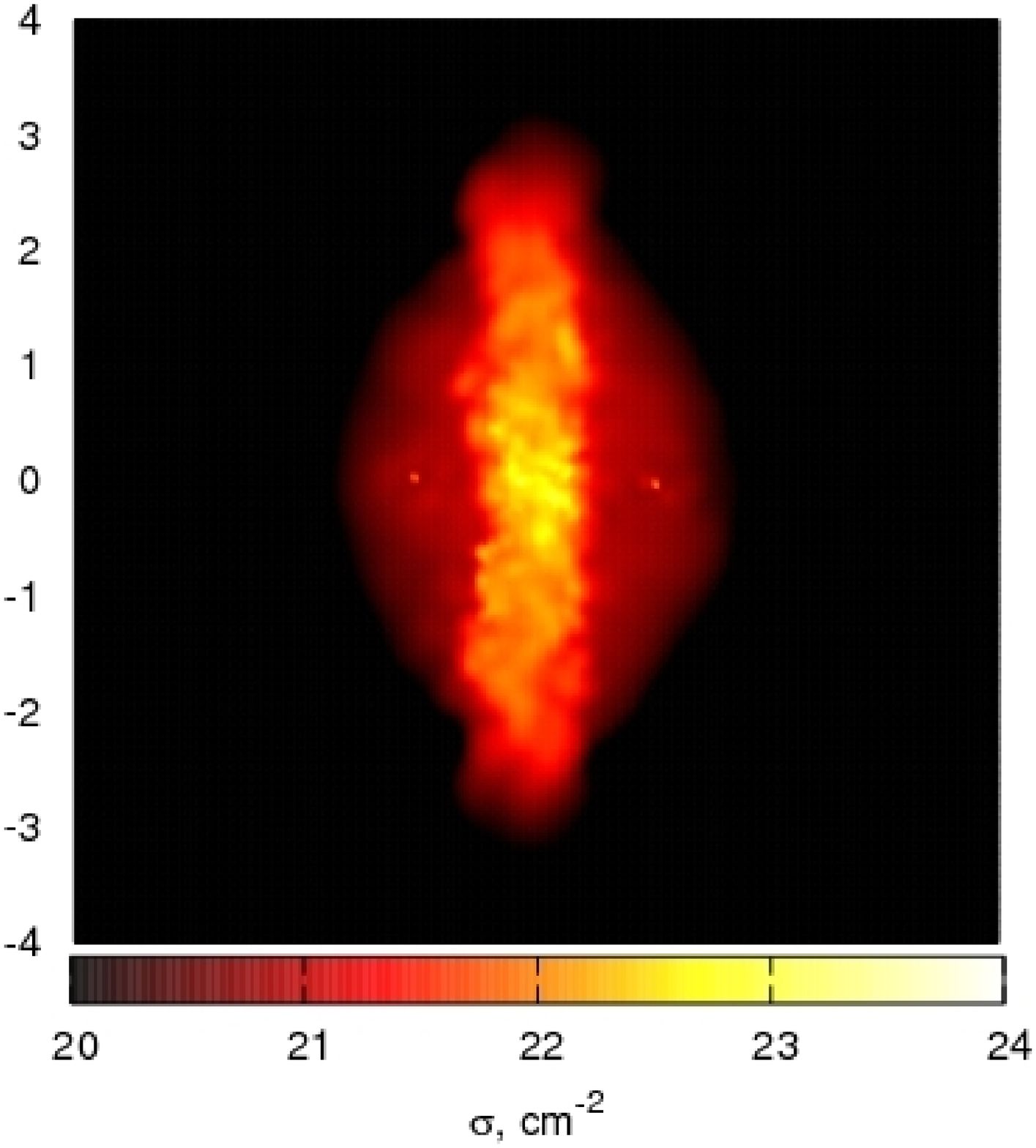} \hfill 
\includegraphics[width=5.2cm]{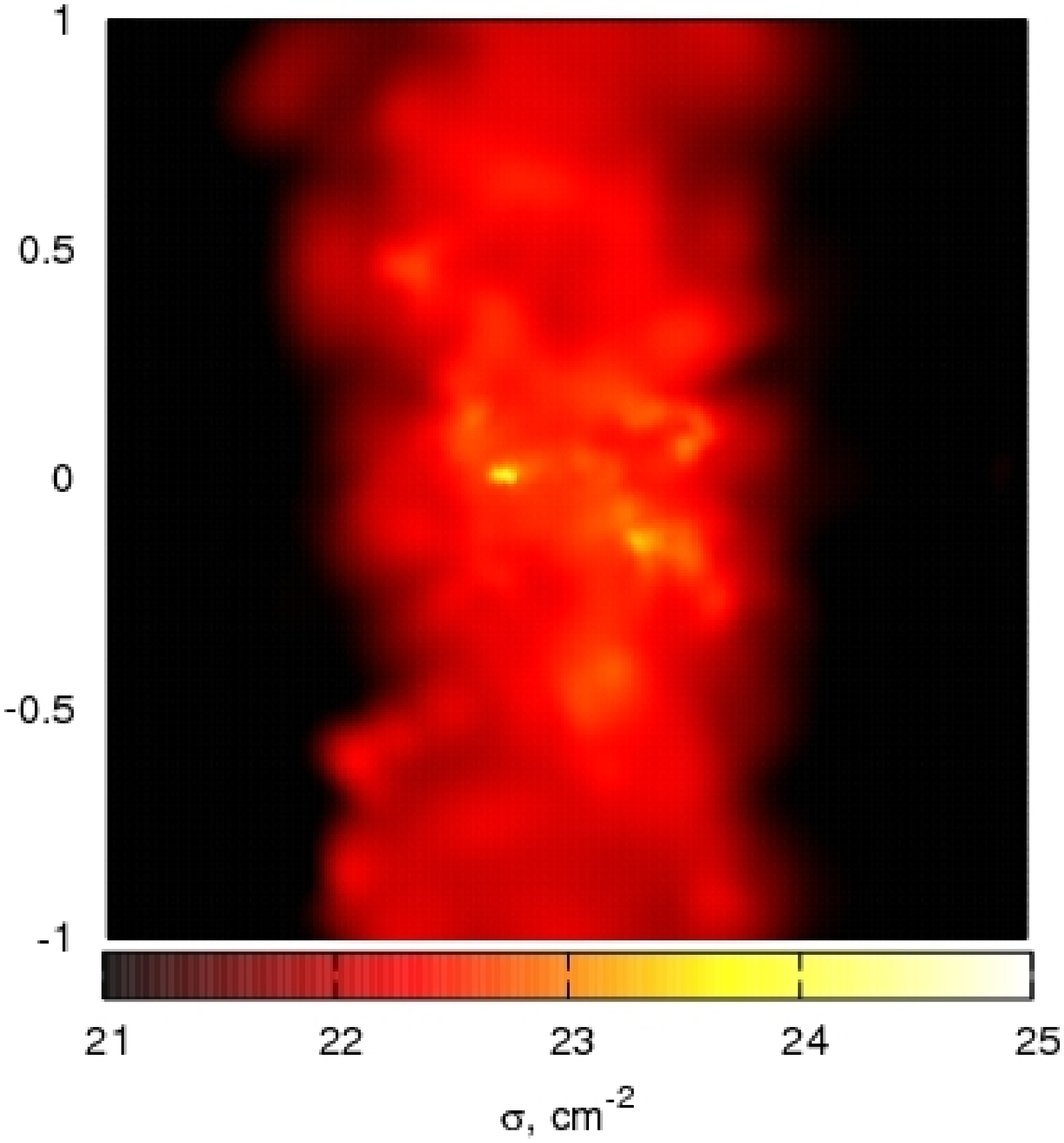} \hfill
\includegraphics[width=5.9cm]{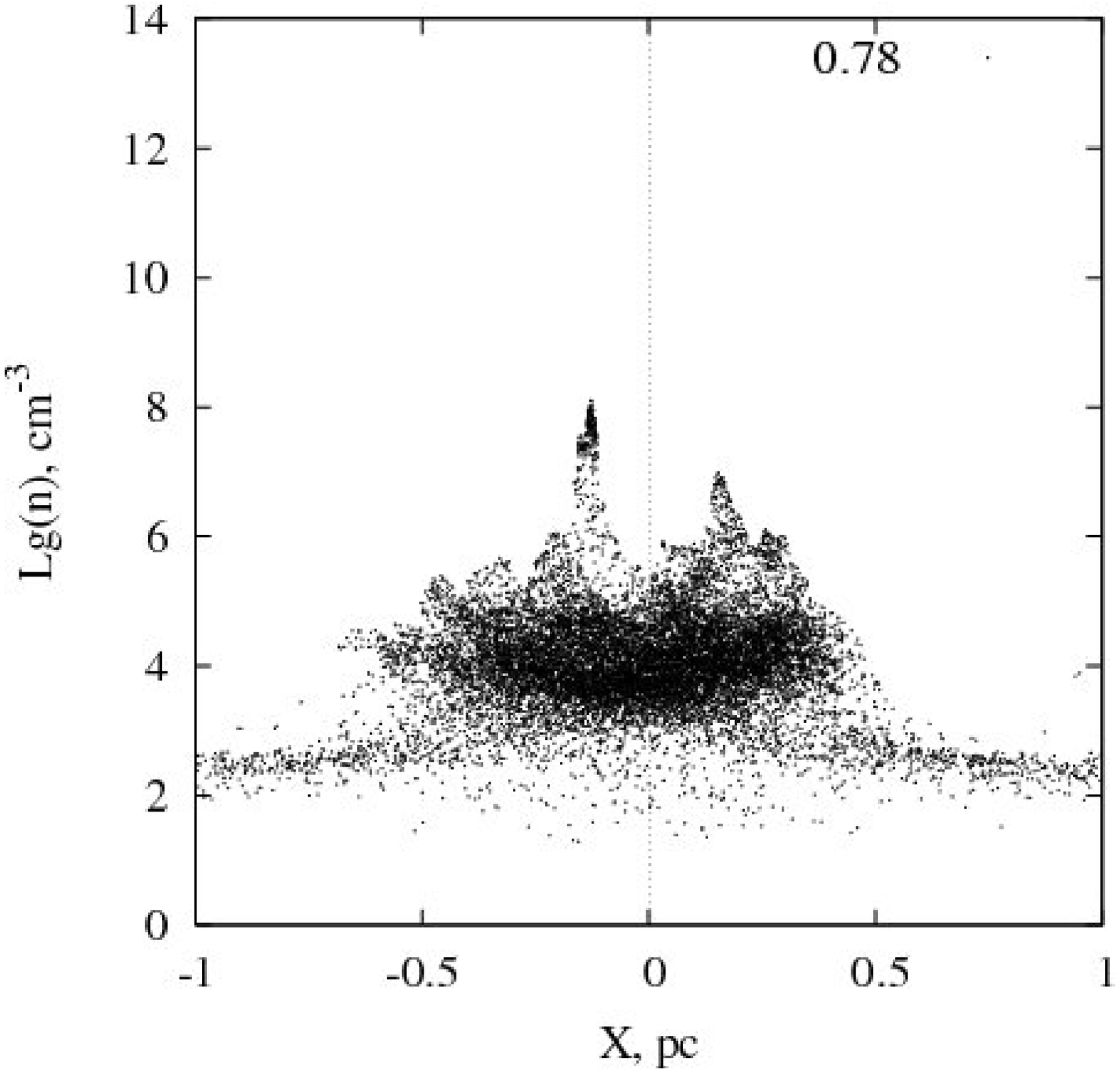} \\
\includegraphics[width=5.0cm]{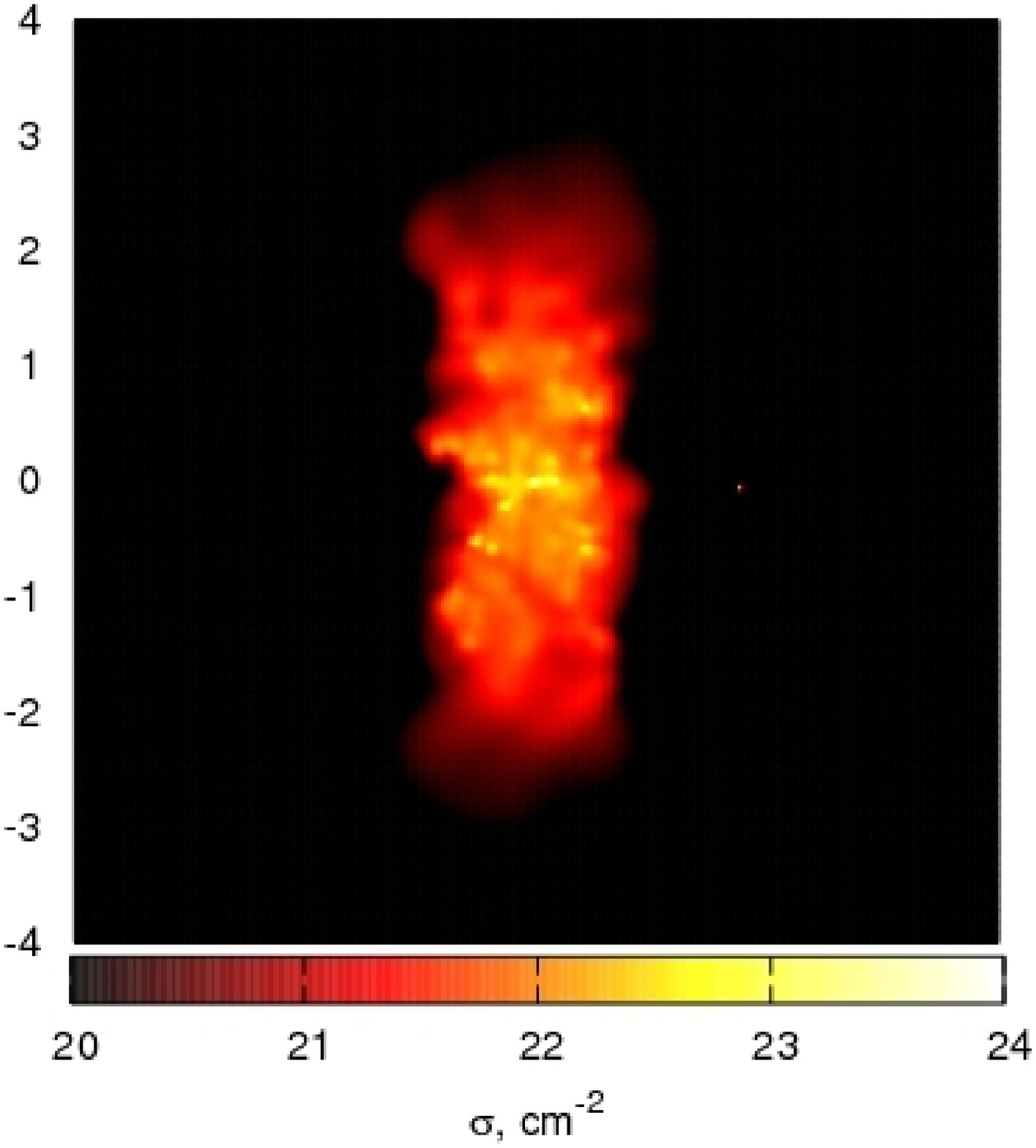} \hfill 
\includegraphics[width=5.2cm]{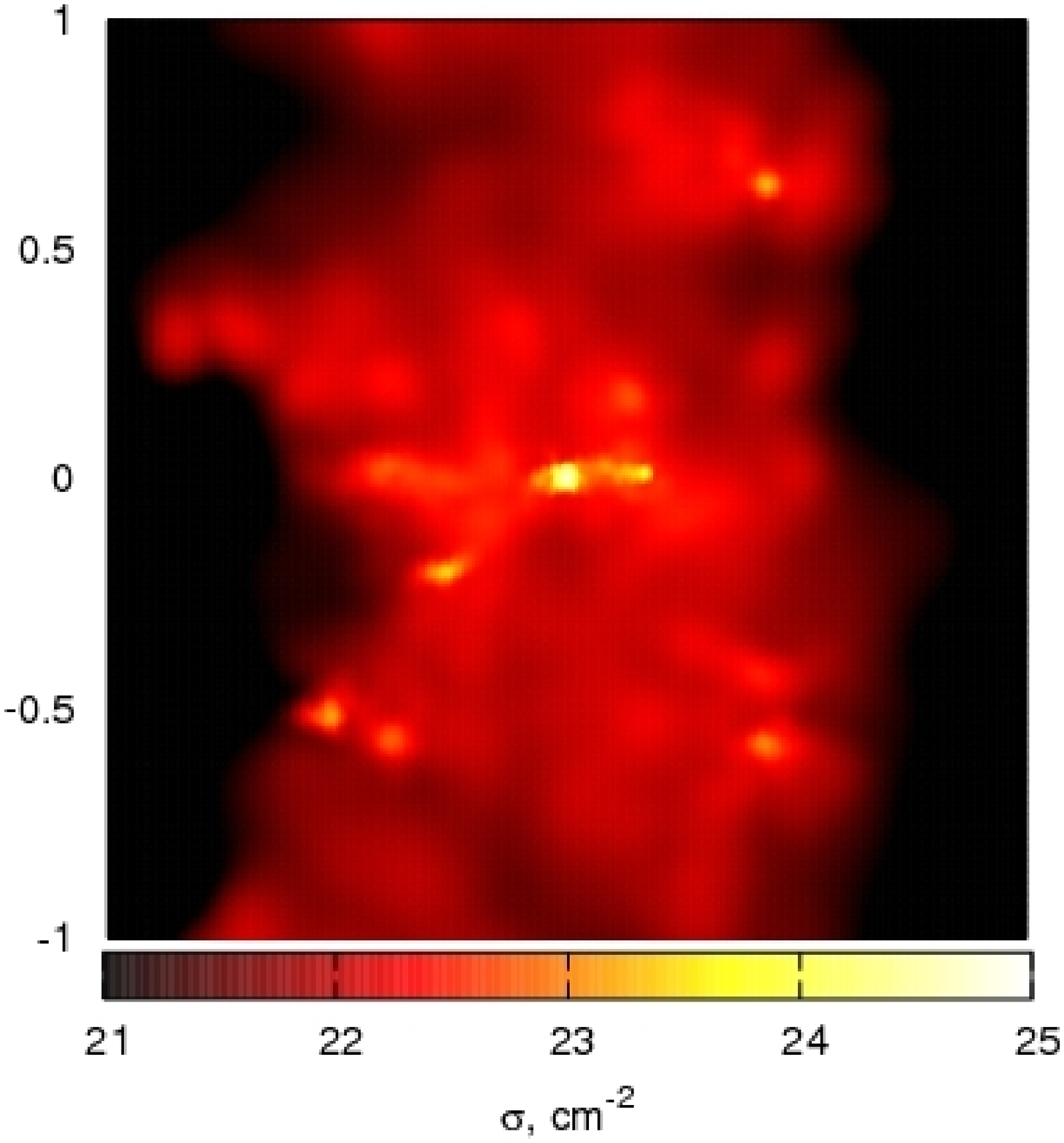} \hfill
\includegraphics[width=5.9cm]{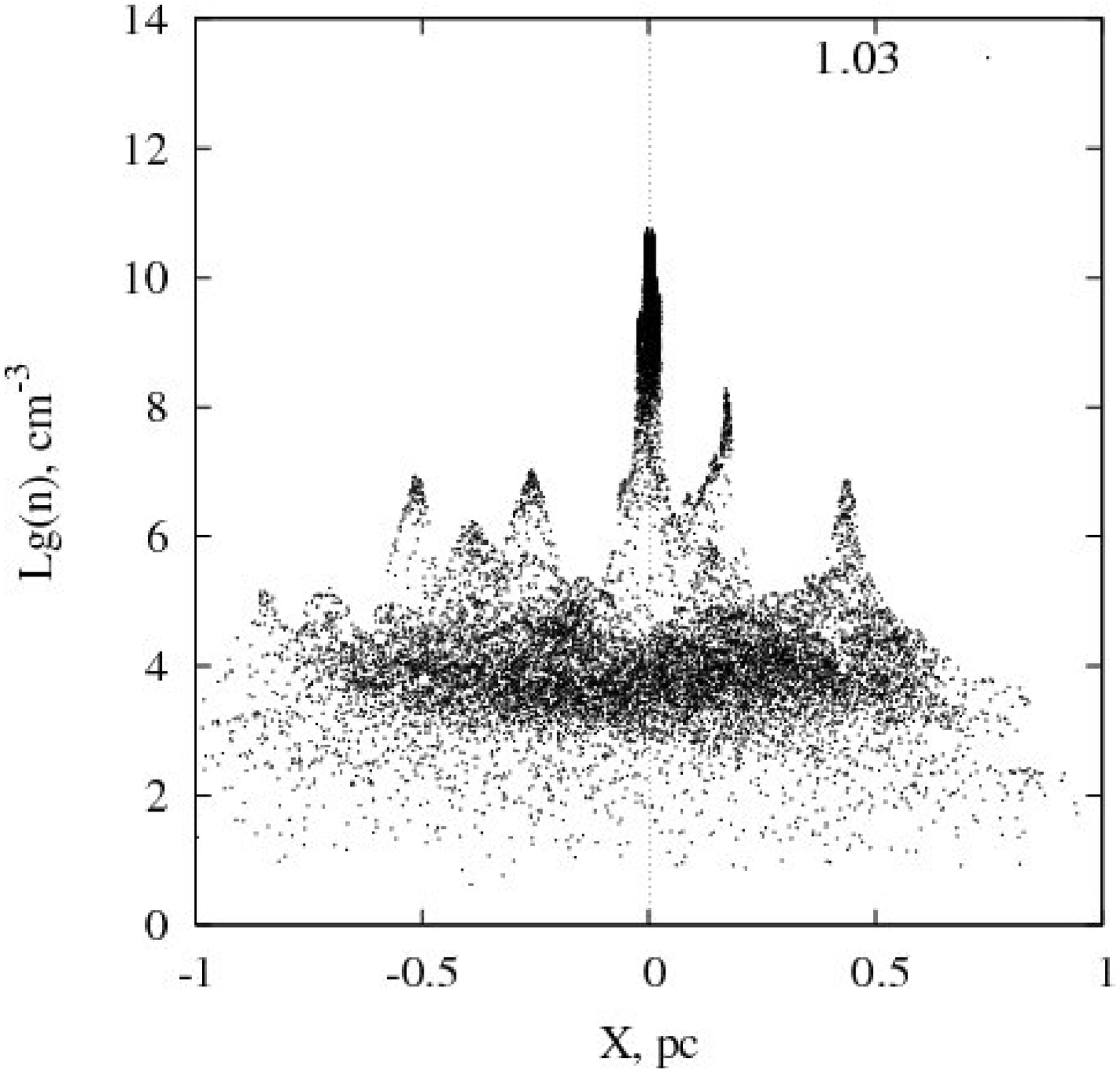}
\caption{Column-density images along $Z$-axis in large and small scales 
         as well as common logarithms of number density \vs{} particles 
         $X$-coordinate in model with $\beta= 0$. The top row is for 
         time $t= 0.78 \; \textrm{Myr}$ and the bottom one -- for 
         $t= 1.03 \; \textrm{Myr}$.}
\protect\label{P:f-1}
\end{figure}

The column-density images in $XY$ plane and particles number densities \vs{} 
$X$-coordinate on two moment are shown on Fig. \ref{P:f-1}

Since the collision was going on between the denser clumps centers, the density in 
shock front had been strong increased and rapid fragmentation there was occurred, 
that yield the first maximum of fragments amount (see Fig. \ref{P:fragments}). 
Next, this fragments began just as rapid merging one with another, giving a few 
massive central objects (minimum of fragments amount on the same plot). Meanwhile, 
the rest gas accreting on them, had been compressed and began intensively fragment 
too, that give the second peak of fragments amount.

Finally, most of gas ($>70 \,\%$) fell onto the flatten dense compact core of mass 
$2921.5 \;\Msun$ and number density up to $\scient{6.7}{11} \; \textrm{cm$^{-3}$}$. 
There are two non-coplanar discs with masses $537$ and $431.5 \;\Msun$ around this 
core. The external ring have three small  subfragments ($\lesssim 20 \;\Msun$) and 
entire system are embedded in thin extensive shell with mass $20 \;\Msun$.

The fact of forming of rotational system in head-on case may be explained by 
features of initial model building (see sec. \ref{S:InitConf}) and therefore there 
is a small impact parameter close to $0.01 \;\textrm{pc}$.

In calculations with half particles number the main picture remain similar to 
above, but the central body had a one single semi-destroyed most massive accretion 
disk abounded it. This "disk" also have one subfragment and entire system was 
embedded in thin shell.

\underline{\textbf{\textsl{Collision with $\beta=0.2$}}} \hspace{7mm}
At such small impact parameter clumps' centers, even if not collided, but passed 
close enough one from another and collision finished, as above, by forming of 
one massive zone of fragmentation 
\myfootnote{\textbf{"zone of fragmentation"} or \textbf{"condensing zone"}}
in center (the regions in which fragments are 
formed by groups) and extent, mainly unfragmented diffusive big two-arm spiral 
with spread in $\simi 30 \;\textrm{pc}$ and thickness from $\simi 3 \;\textrm{pc}$ 
in center to $\simi 10 \;\textrm{pc}$ on periphery. In contrast to center, each 
arm consist from gas of its "own" clump.

\begin{figure}[Htbp]
\centering
\includegraphics[width=5.0cm]{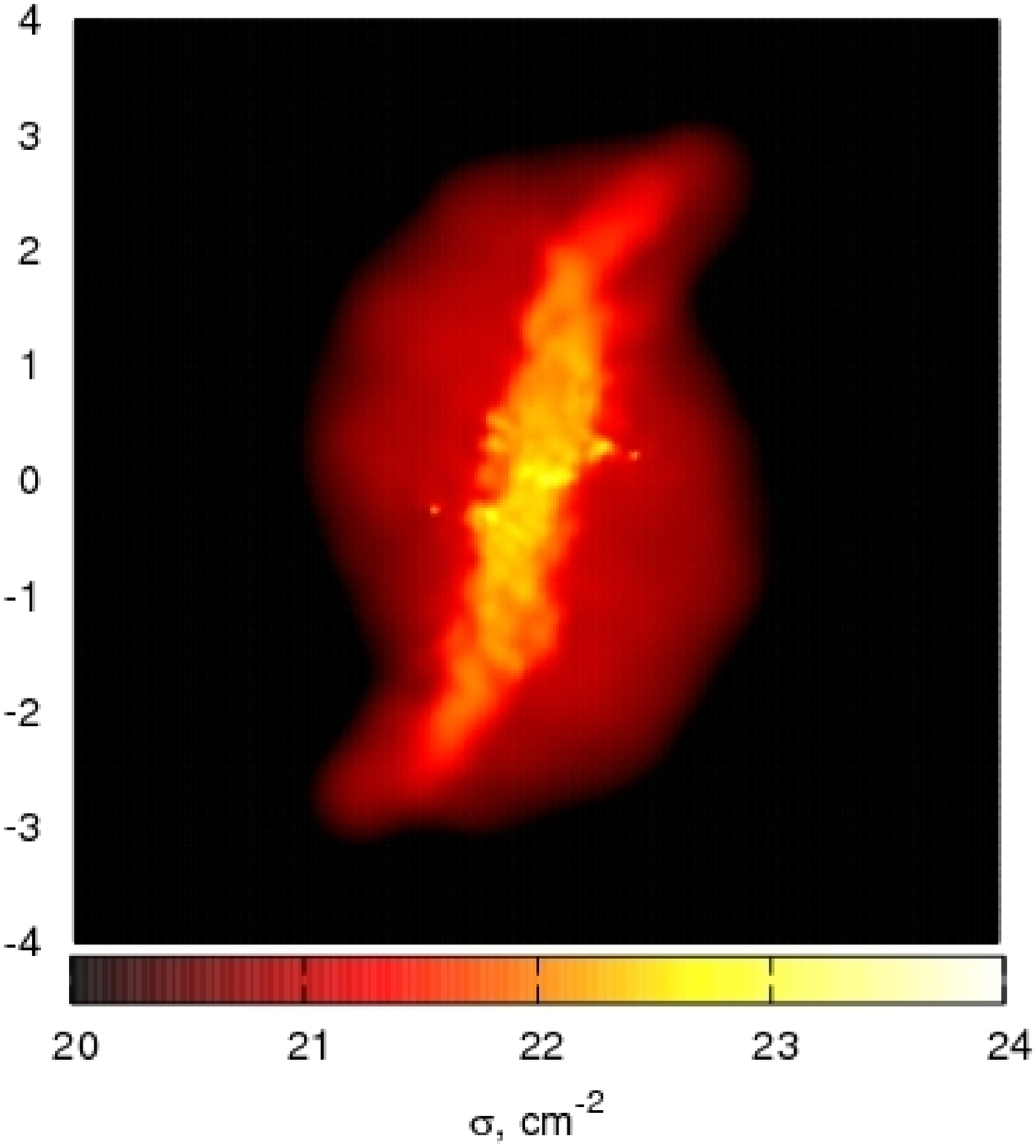} \hfill 
\includegraphics[width=5.2cm]{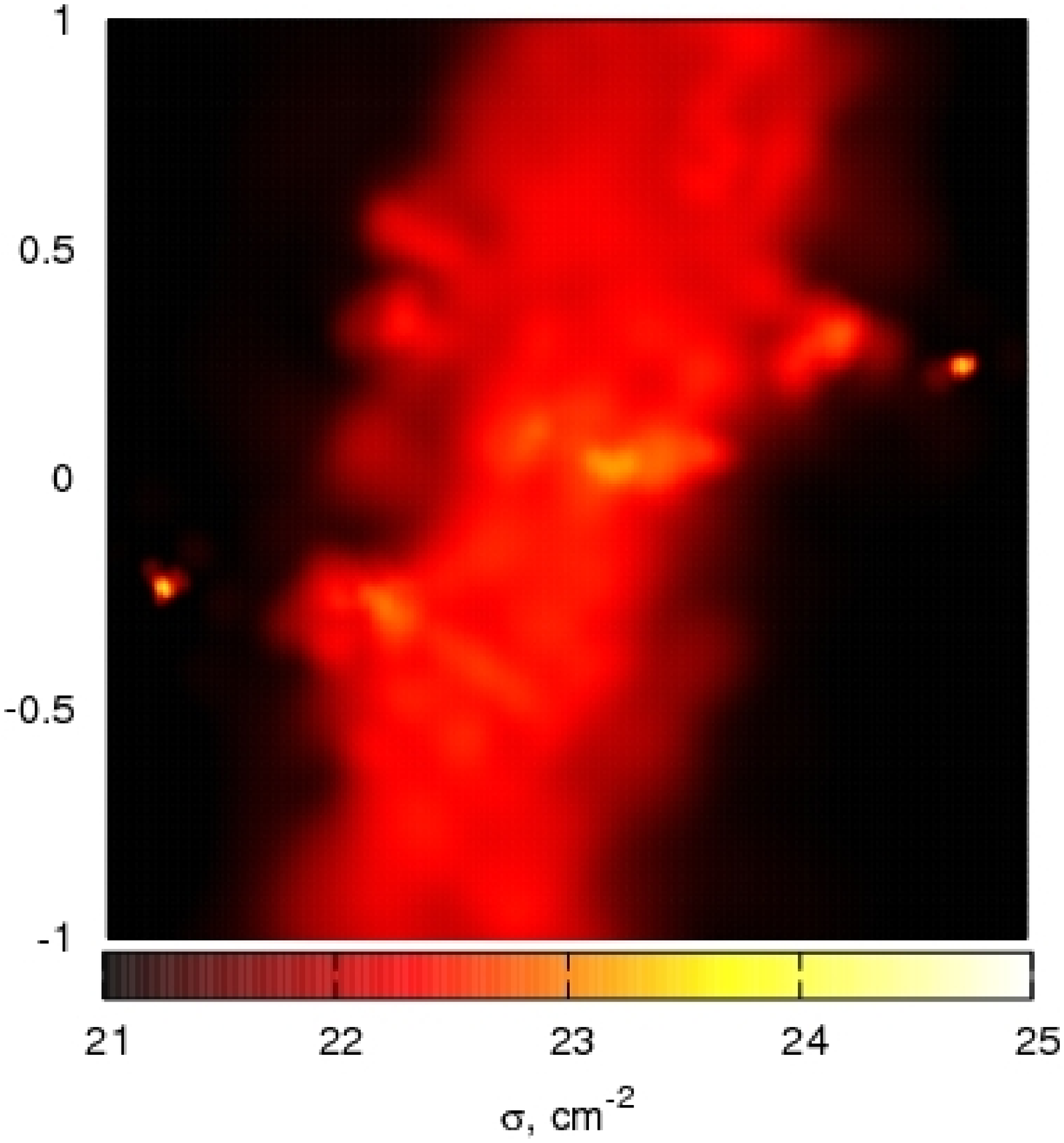} \hfill
\includegraphics[width=5.9cm]{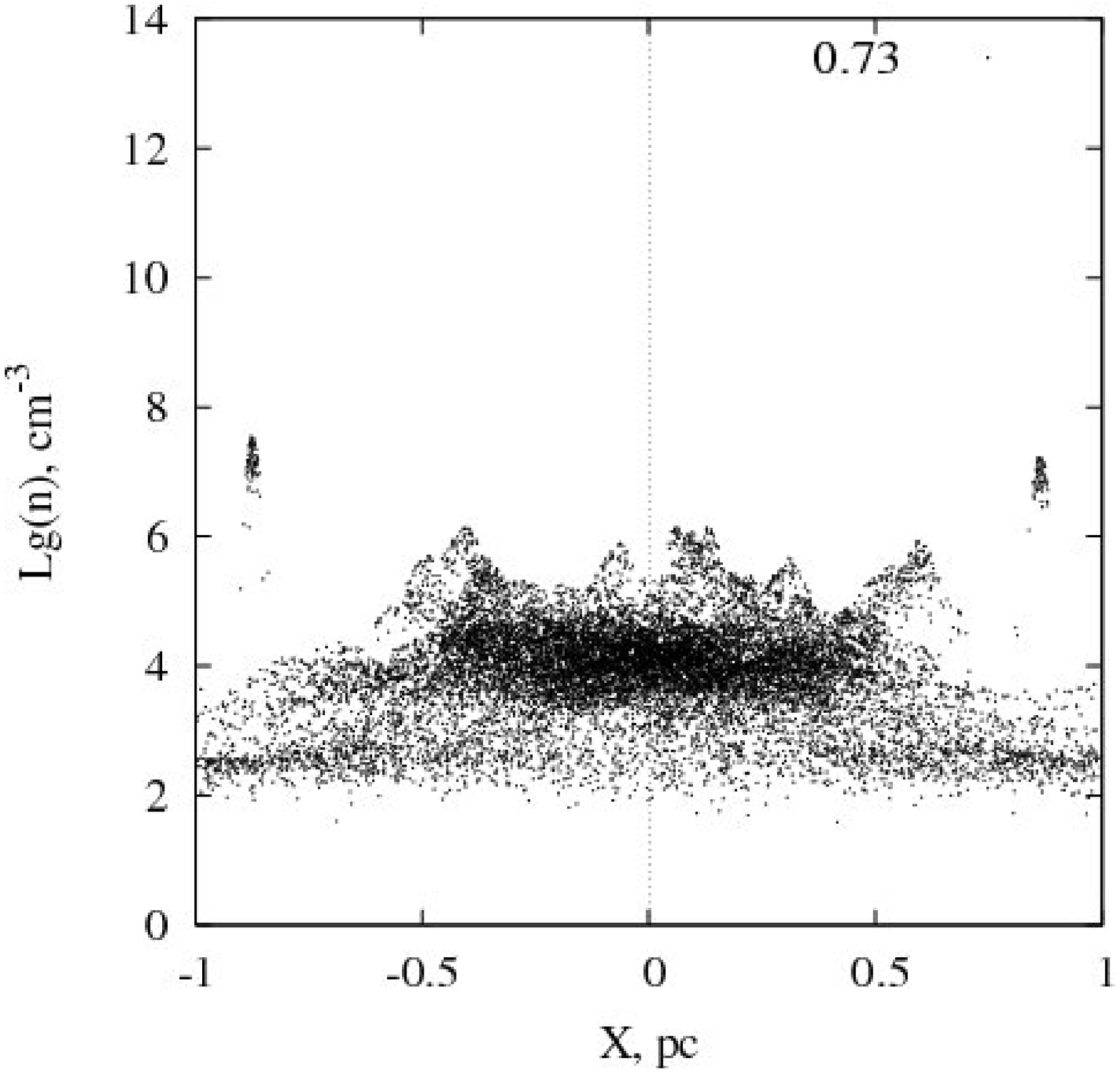} \\
\includegraphics[width=5.0cm]{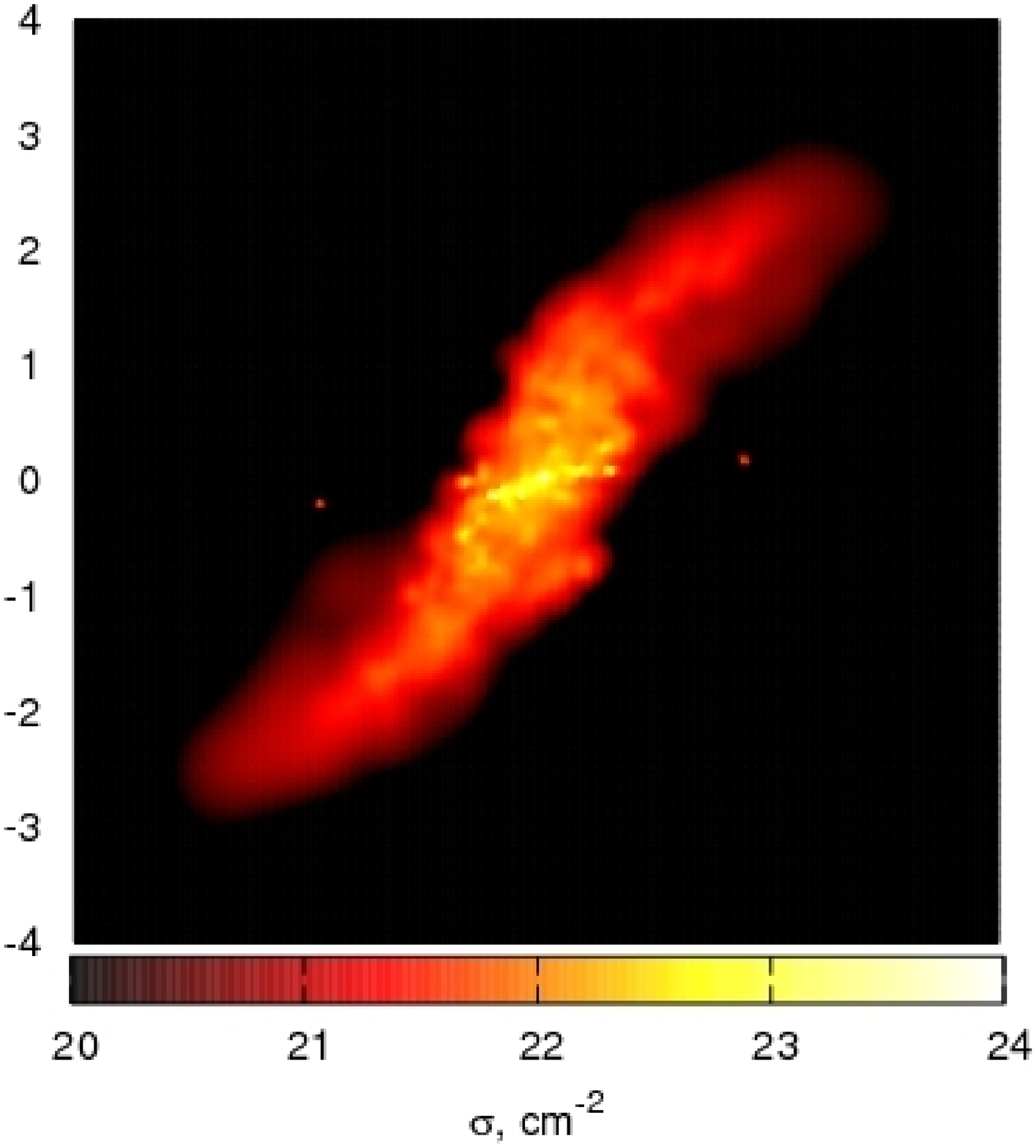} \hfill 
\includegraphics[width=5.2cm]{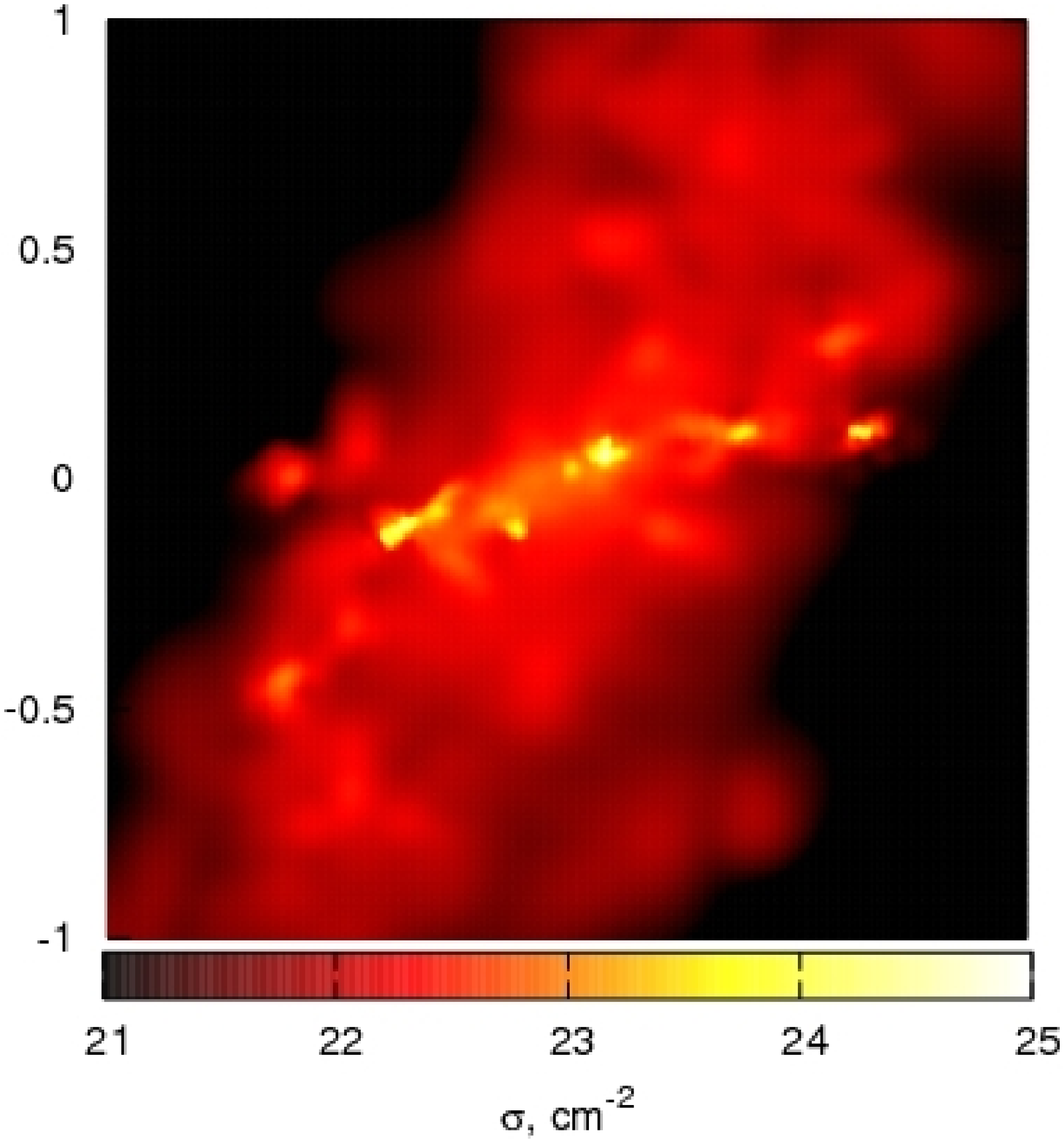} \hfill
\includegraphics[width=5.9cm]{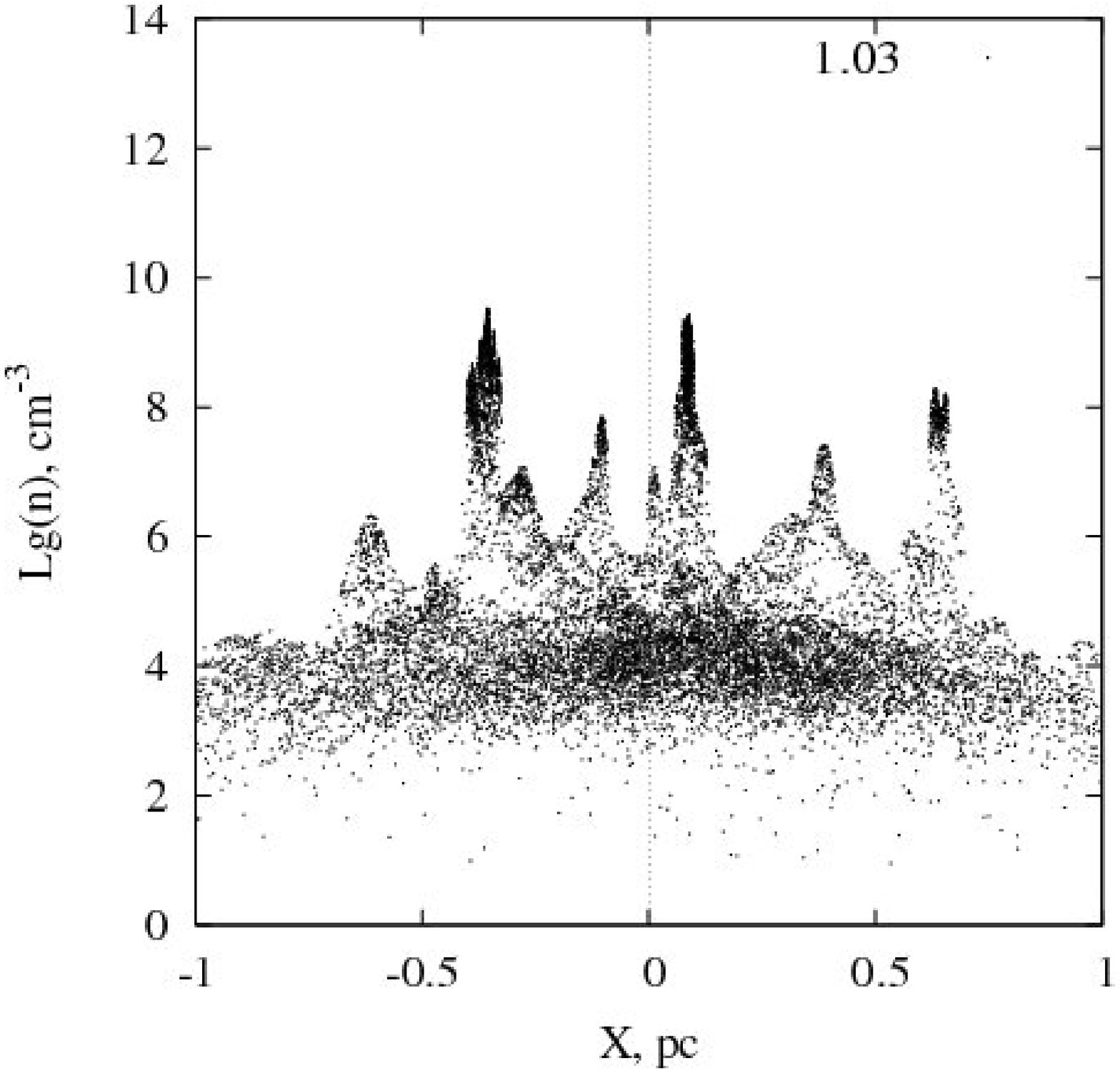}
\caption{Column-density images with $X-\lg(n)$ plots for $\beta= 0.2$.
         The top row is for time $t= 0.73 \; \textrm{Myr}$ and the bottom 
         one -- for $t= 1.03 \; \textrm{Myr}$.
         See notes on Fig. \ref{P:f-1}}
\protect\label{P:f-2}
\end{figure}

The some moment of evolution are illustrated in Fig. \ref{P:f-2}. But, at the 
end of simulated we have the massive flatten core with mass in $1141 
\;\Msun$ and number density $\lesssim \scient{1.7}{10} \;\textrm{cm$^{-3}$}$ 
in center and its disc-shell by mass $1763.5 \;\Msun$, which oneself have 
three petty subfragments ($10-15 \;\Msun$). They are surrounded by internal 
two-armed spiral with size close to $1.5-2 \;\textrm{pc}$, that have four 
fragments ($164.25$, $24$, $16.75$ and $12 \;\Msun$). Rest fragments with masses 
nearly to $20 \;\Msun$ are detected in both arms of great spiral.

In simulations with $8000$ particles the external spiral was practically 
unfragmented with the exception of one fragment with mass $22.5 \Msun$. 

The core looks like a one-armed spiral of $\simi 0.75 \;\textrm{pc}$ 
and three satellites with masses in a tens solar masses.

\underline{\textbf{\textsl{Collision with $\beta=0.5$}}} \hspace{7mm}
At $\beta = 0.50$ central clumps' regions, as before, passed through shock front, 
has been formed and where at $\simi 0.6 \;\textrm{Myr}$ appeared the first 
fragments with good mixed gas. Rest material, mainly from outer layers of clumps, 
go past far from "place of collision" and form large two-armed spiral structure 
about of $40 \;\textrm{pc}$. The next fragments generation had been formed in 
outer sided of both arm after $\simi 1 \;\textrm{Myr}$ since extensively 
fragments formation in shock front.

Now collision gave 4-5 big zones of fragmentation, instead of only one such zone 
in two previous cases. They are the close pair in center, one single fragment 
about them and two less massive condensations on the both ends of spiral arms.

\begin{figure}[Htbp]
\vspace{-2cm}
\centering
\includegraphics[width=5.0cm]{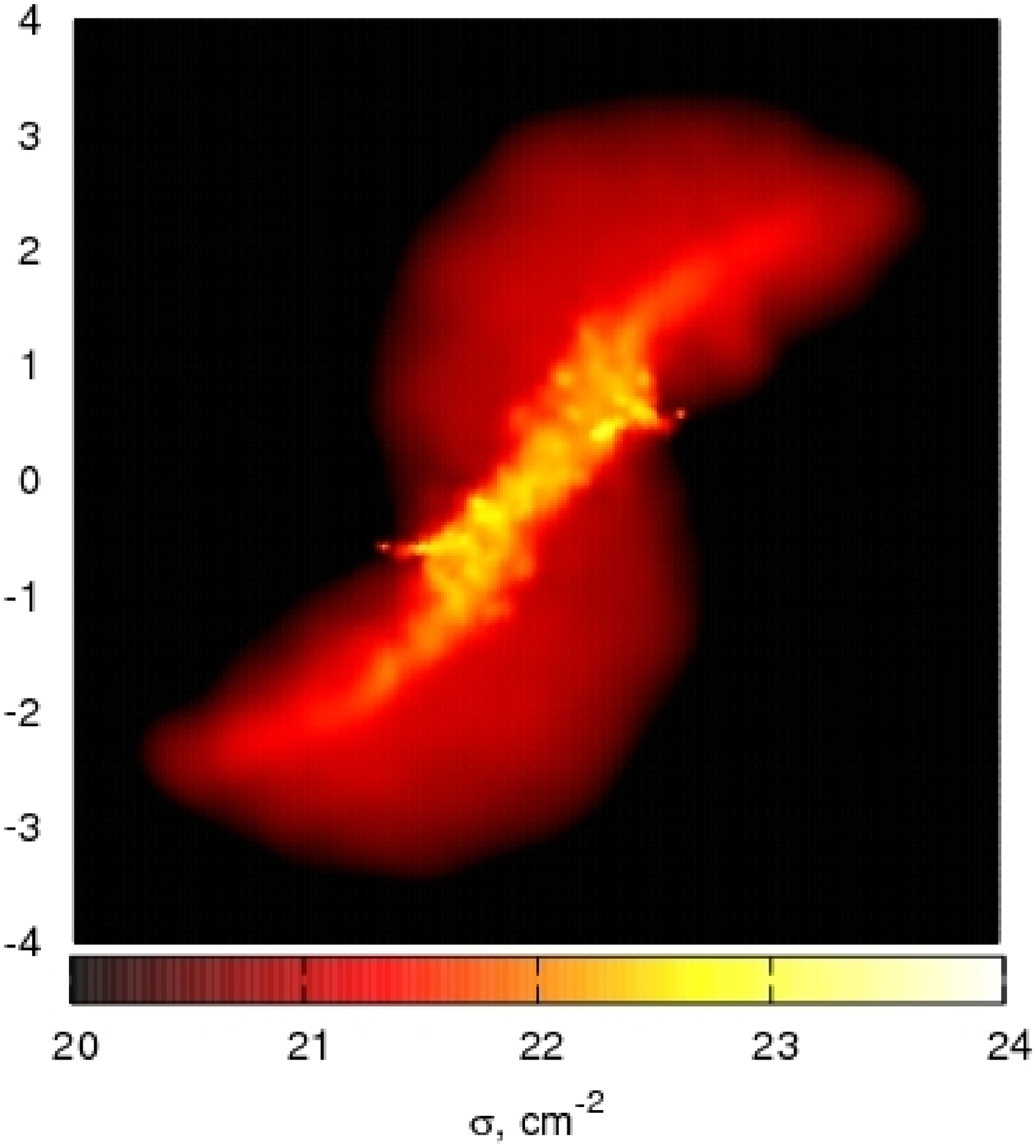} \hfill 
\includegraphics[width=5.2cm]{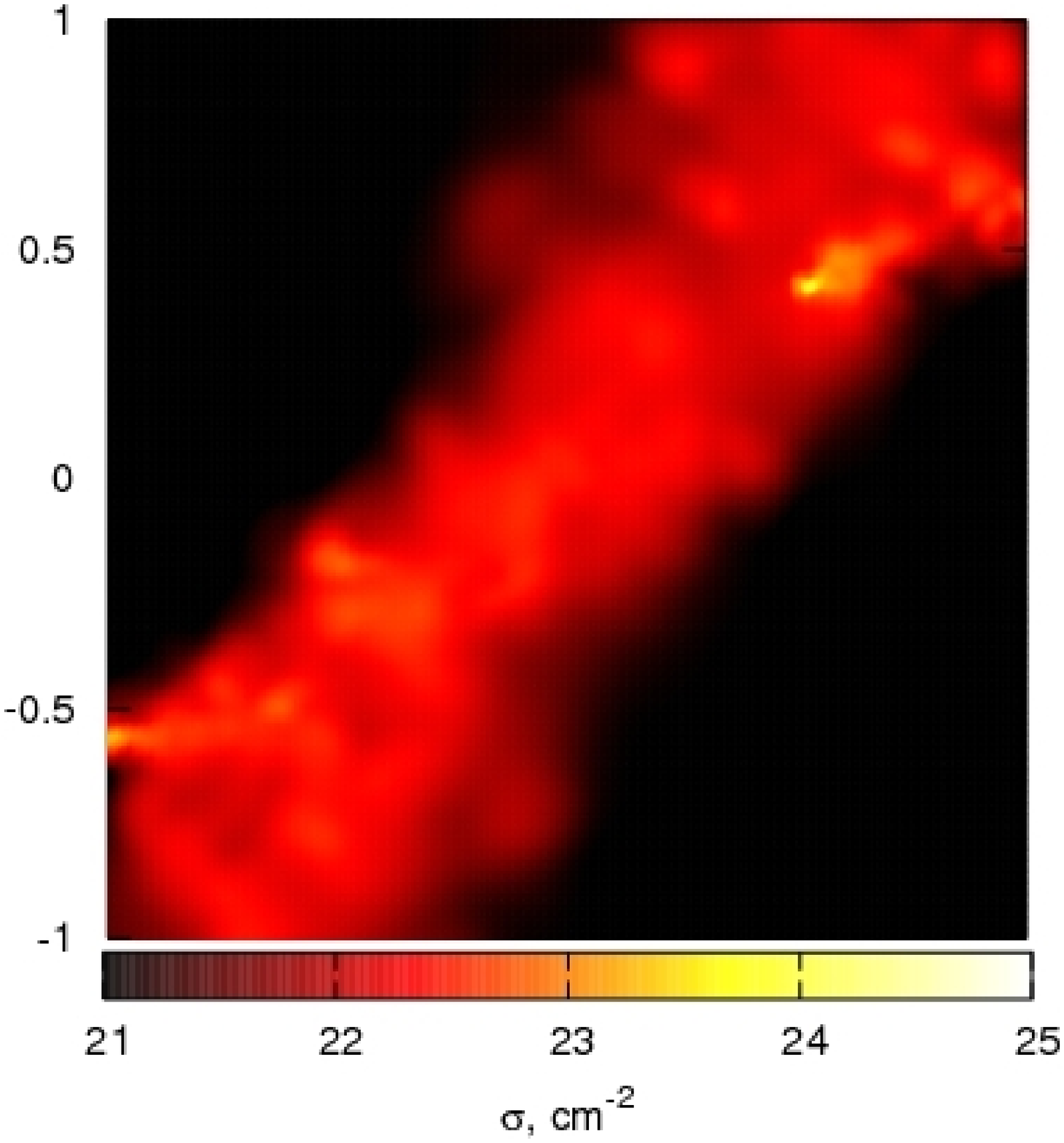} \hfill
\includegraphics[width=5.9cm]{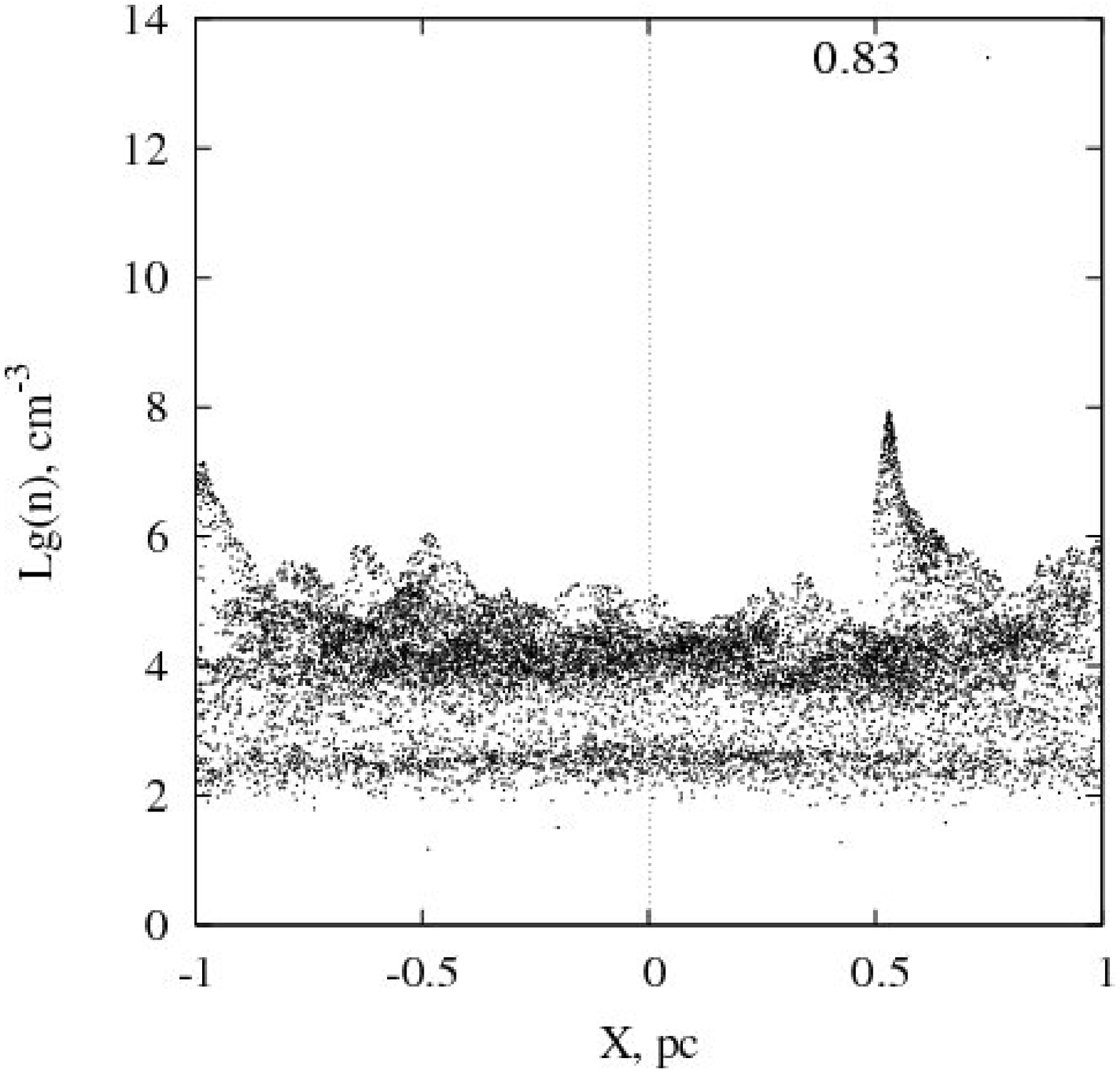} \\
\includegraphics[width=5.0cm]{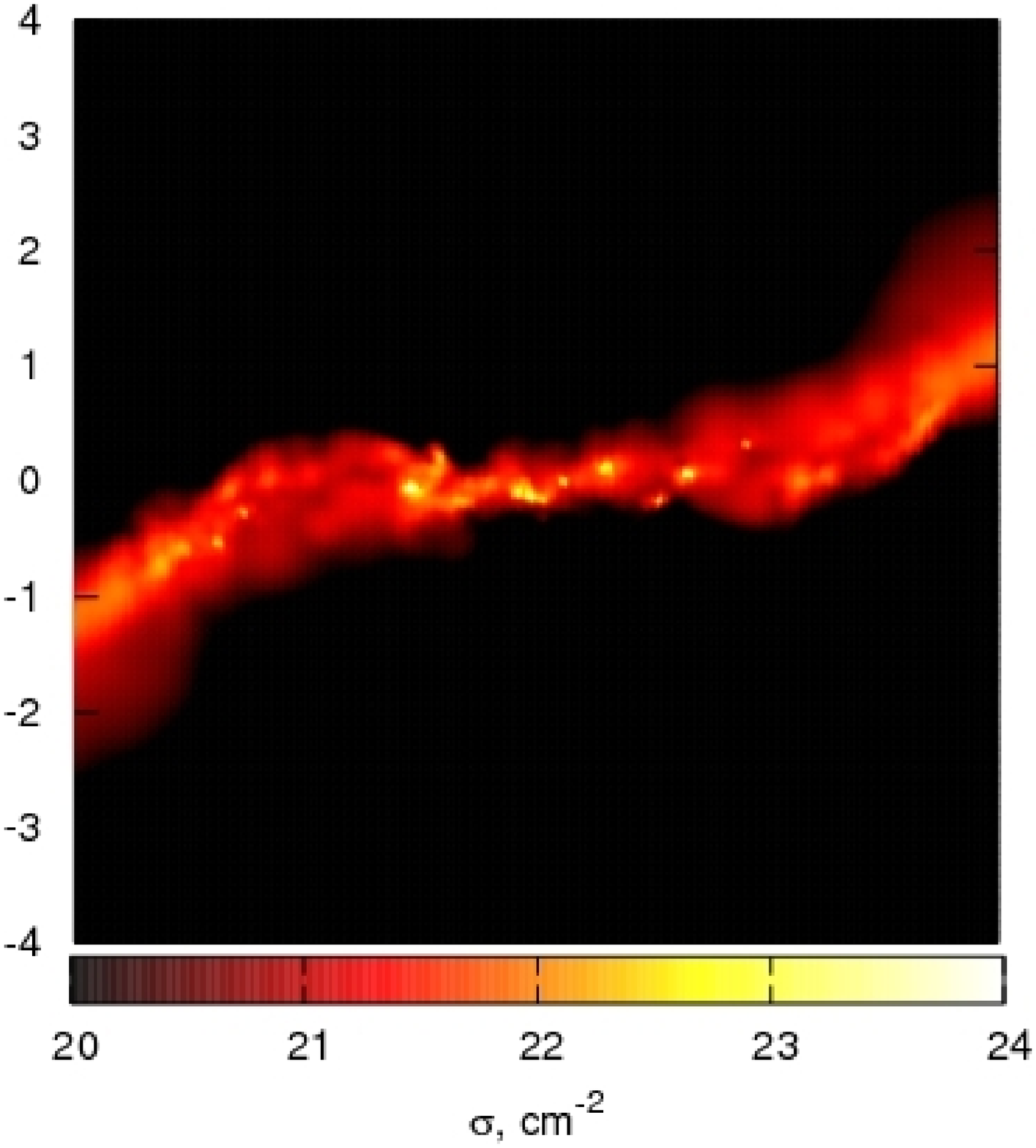} \hfill 
\includegraphics[width=5.2cm]{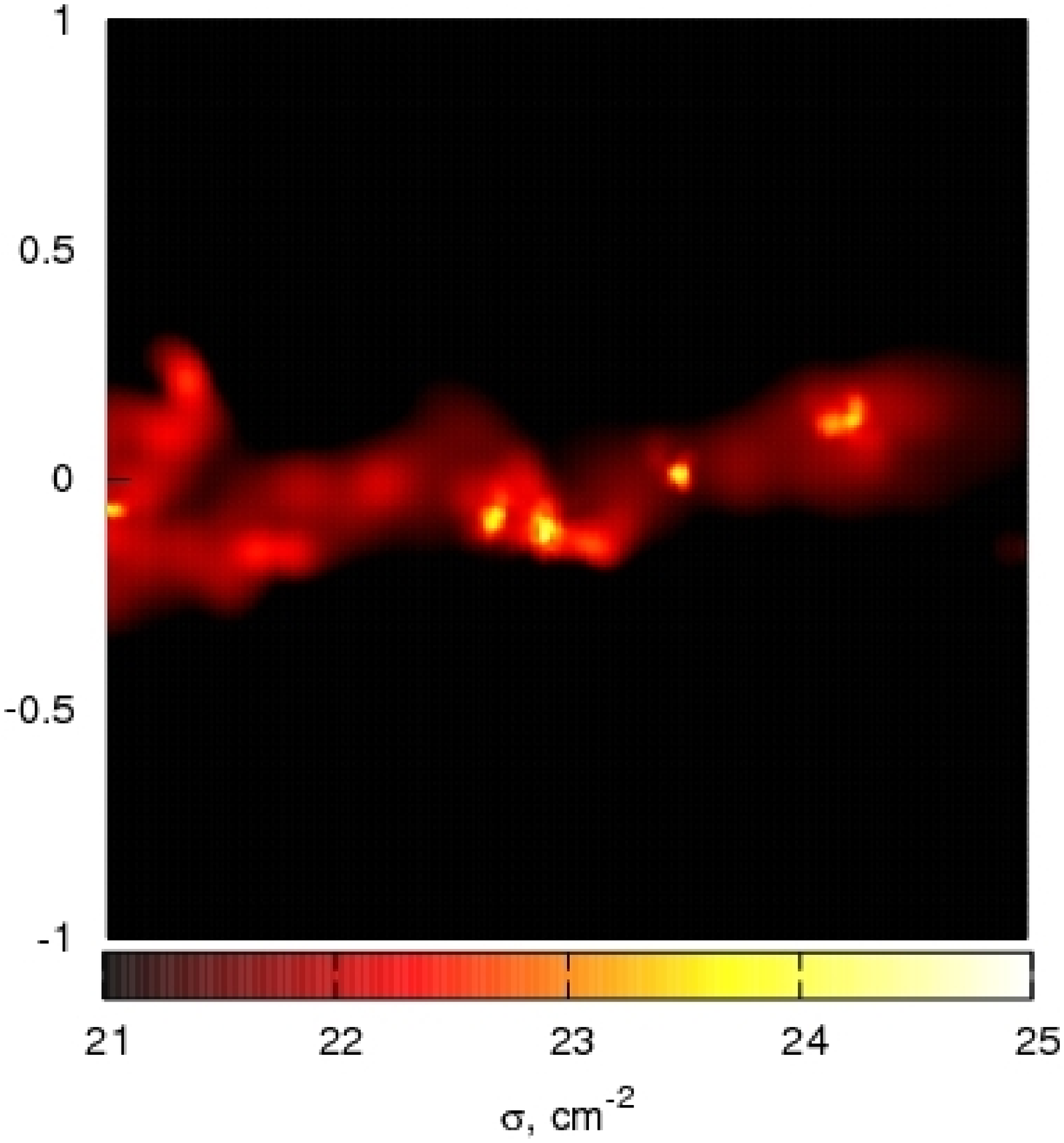} \hfill
\includegraphics[width=5.9cm]{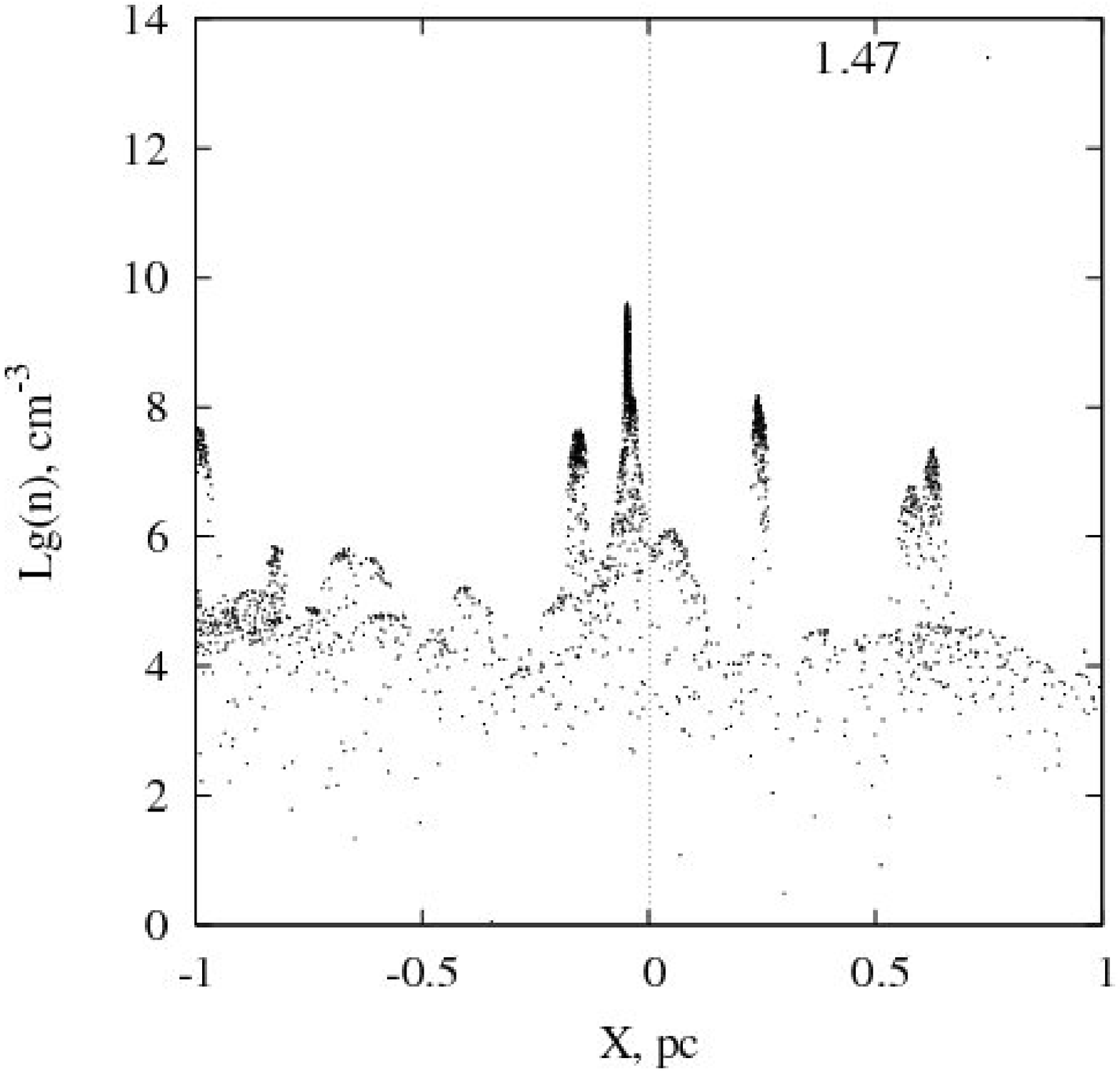}
\caption{Column-density images with $X-\lg(n)$ plots for $\beta= 0.5$.
         The top row is for time $t= 0.83 \; \textrm{Myr}$ and the bottom 
         one -- for $t= 1.47 \; \textrm{Myr}$.
         See notes on Fig. \ref{P:f-1}}
\protect\label{P:f-3}
\end{figure}

The column-density images and $X$ \vs{} $Lg(n)$ plots at $0.83$ and $1.47 \
;\textrm{Myr}$ are shown on Fig. \ref{P:f-3}.

The central pair is two massive flatten cores of masses $563.25$ and $475.5 
\;\Msun$ with number densities beyond $10^9 \;\textrm{cm$^{-3}$}$, which are 
surrounded by accretion discs of masses $563.75$ and $610 \;\Msun$, correspondingly 
and by order less dense. But, the first core yet have low massive thin flat 
shell ($23.25 \;\Msun$) around, and second one continuously (without space) 
transform to their disc-shell.

In vicinity of central pair there is a third smaller dense fragment with mass
$351.5 \;\Msun$ and number density $\lesssim \scient{2.7}{9} \;\textrm{cm$^{-3}$}$.

There are core $399.75 \;\Msun$, $(n \lesssim \scient{1.5}{9} \;\textrm{cm$^{-3}$})$ 
with double-shell of $73.75 \;\Msun$ on the end of one arm, and another one $382.75 
\;\Msun$, $(n \lesssim \scient{8.9}{8} \;\textrm{cm$^{-3}$})$ with shell in $54.5 
\;\Msun$ on the other side.

As for calculations with half mass resolution, in the first place spiral arms 
are mainly non-fragmented, also, both central core have accretion discs and 
surrounded by thin flat shells.

\begin{figure}[Htbp]
\centering
\includegraphics[width=5.0cm]{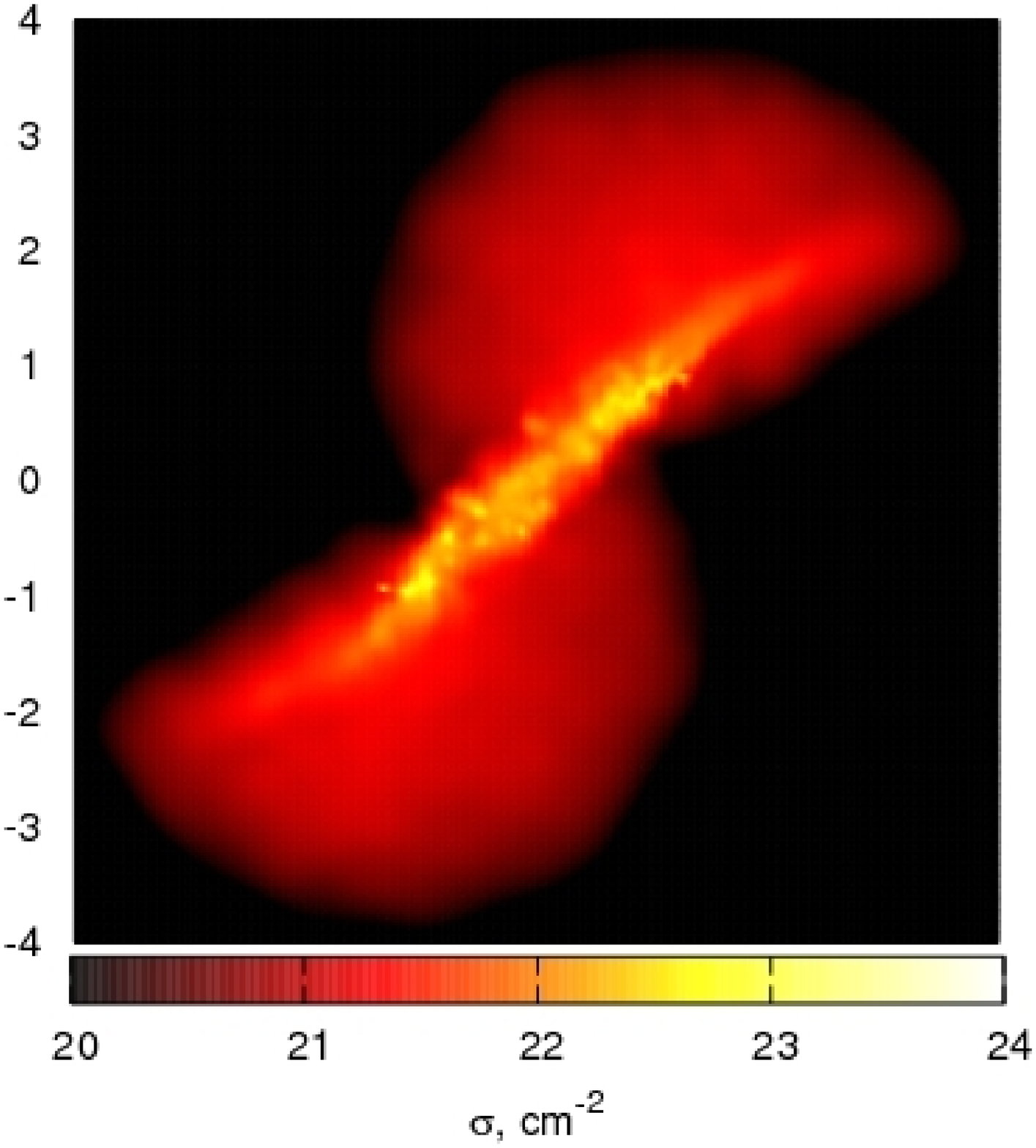} \hfill 
\includegraphics[width=5.2cm]{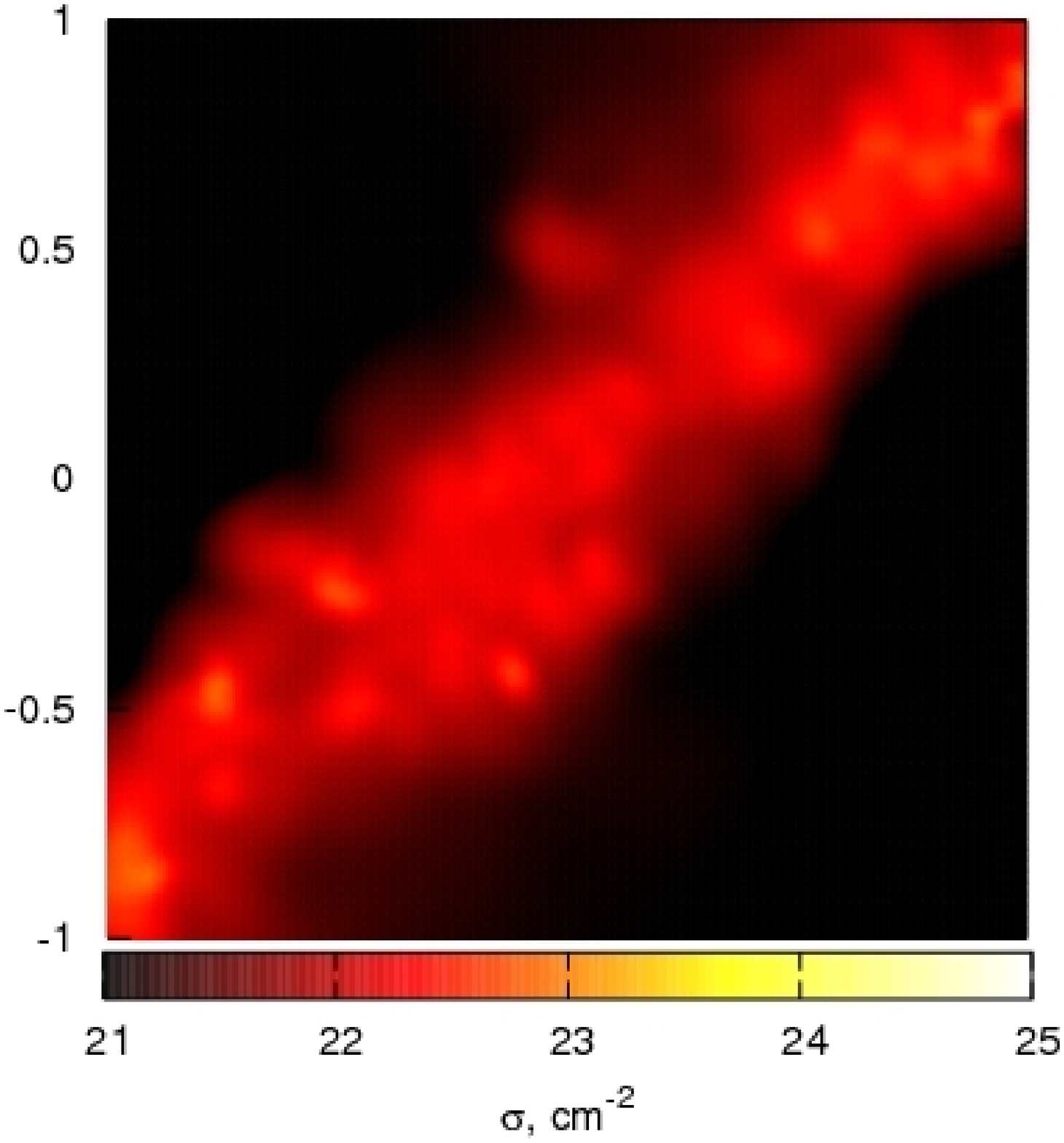} \hfill
\includegraphics[width=5.9cm]{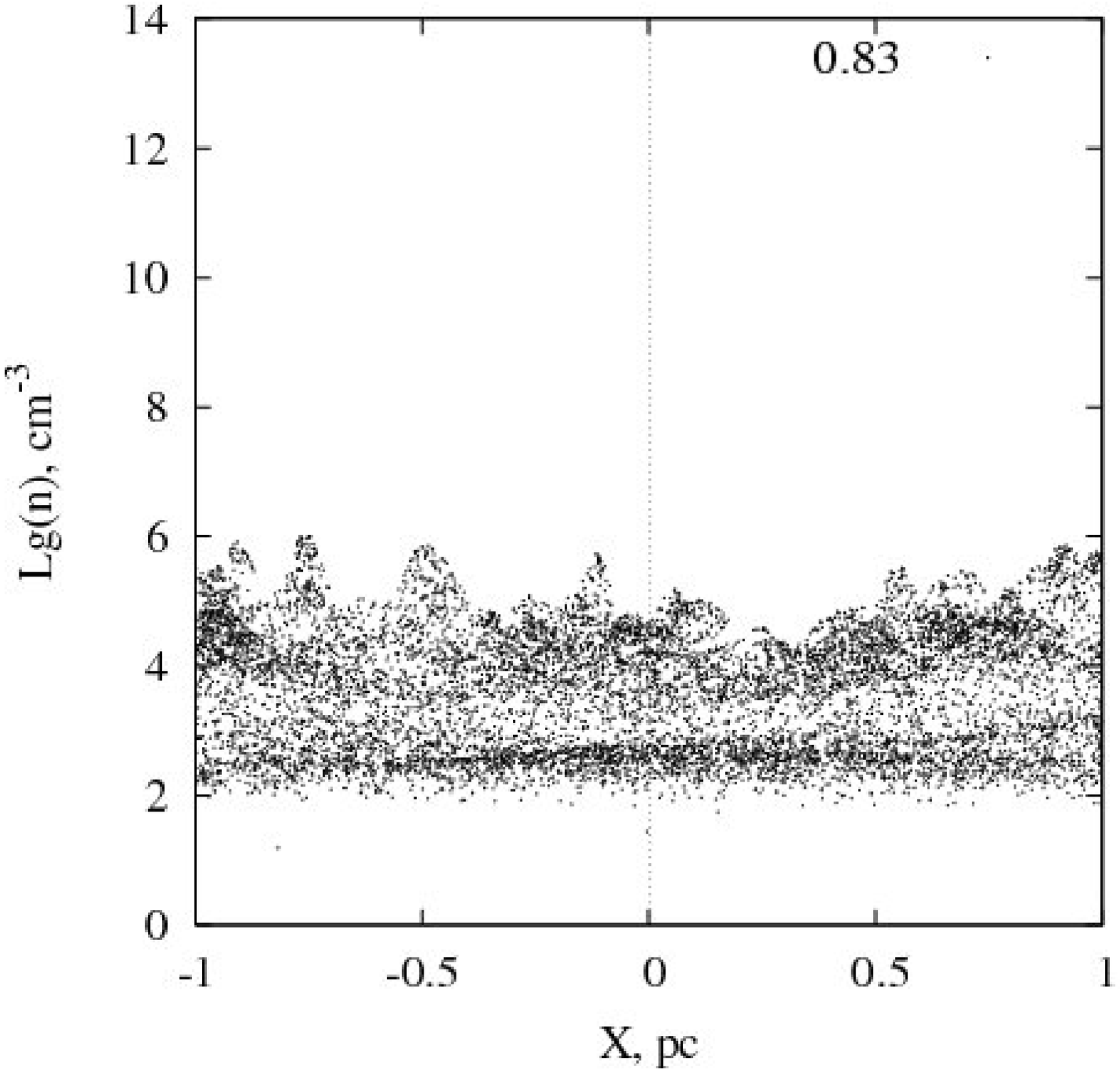} \\
\includegraphics[width=5.0cm]{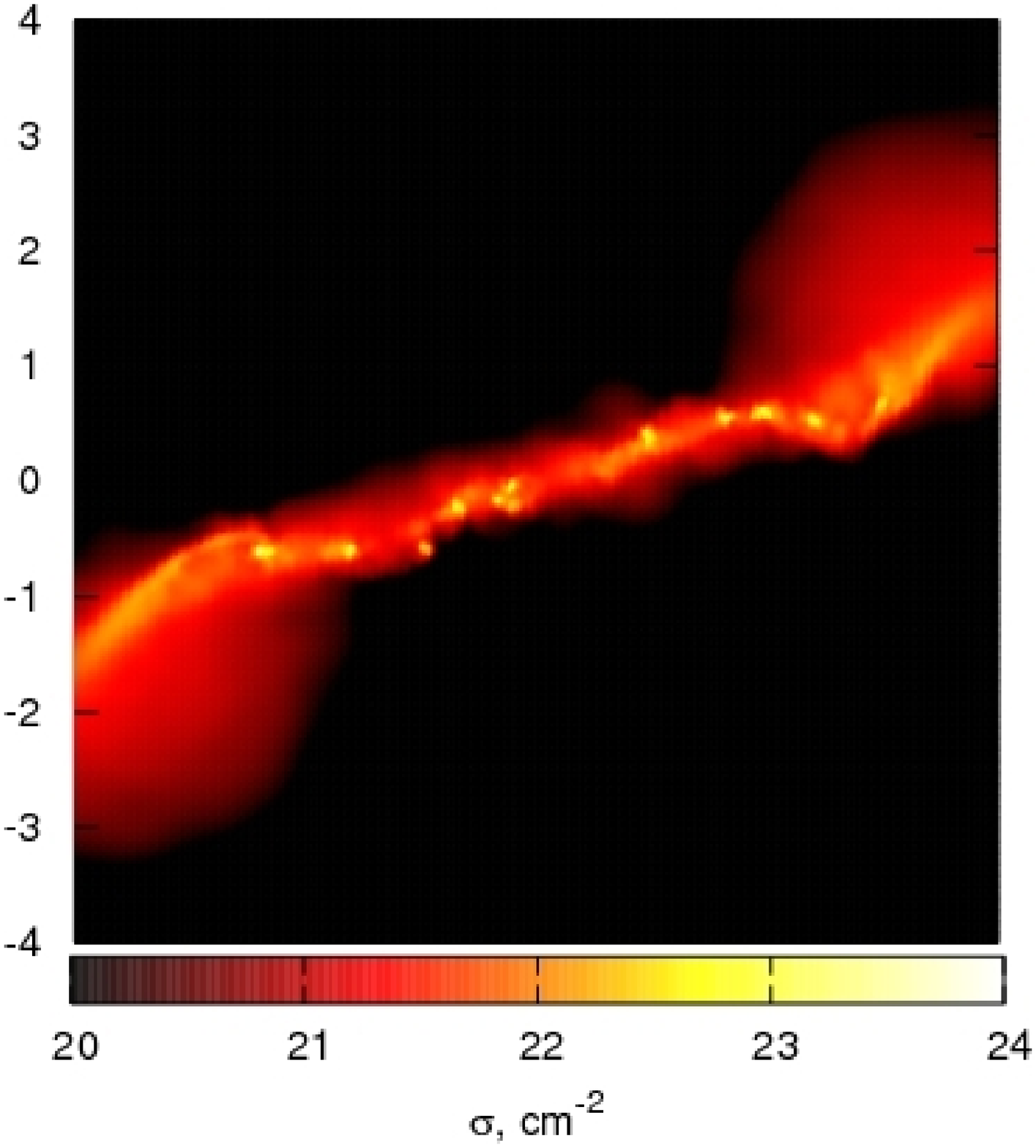} \hfill 
\includegraphics[width=5.2cm]{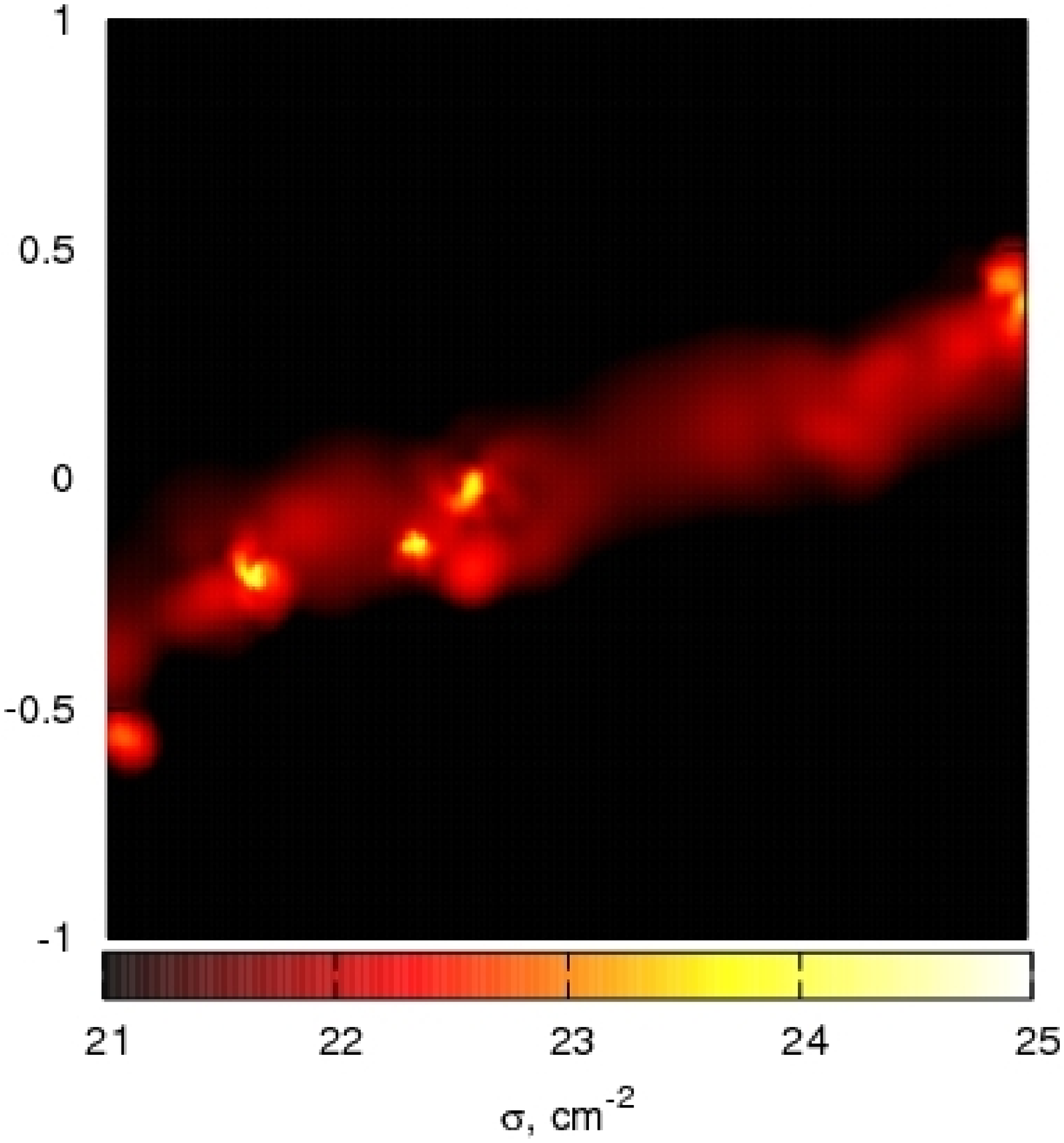} \hfill
\includegraphics[width=5.9cm]{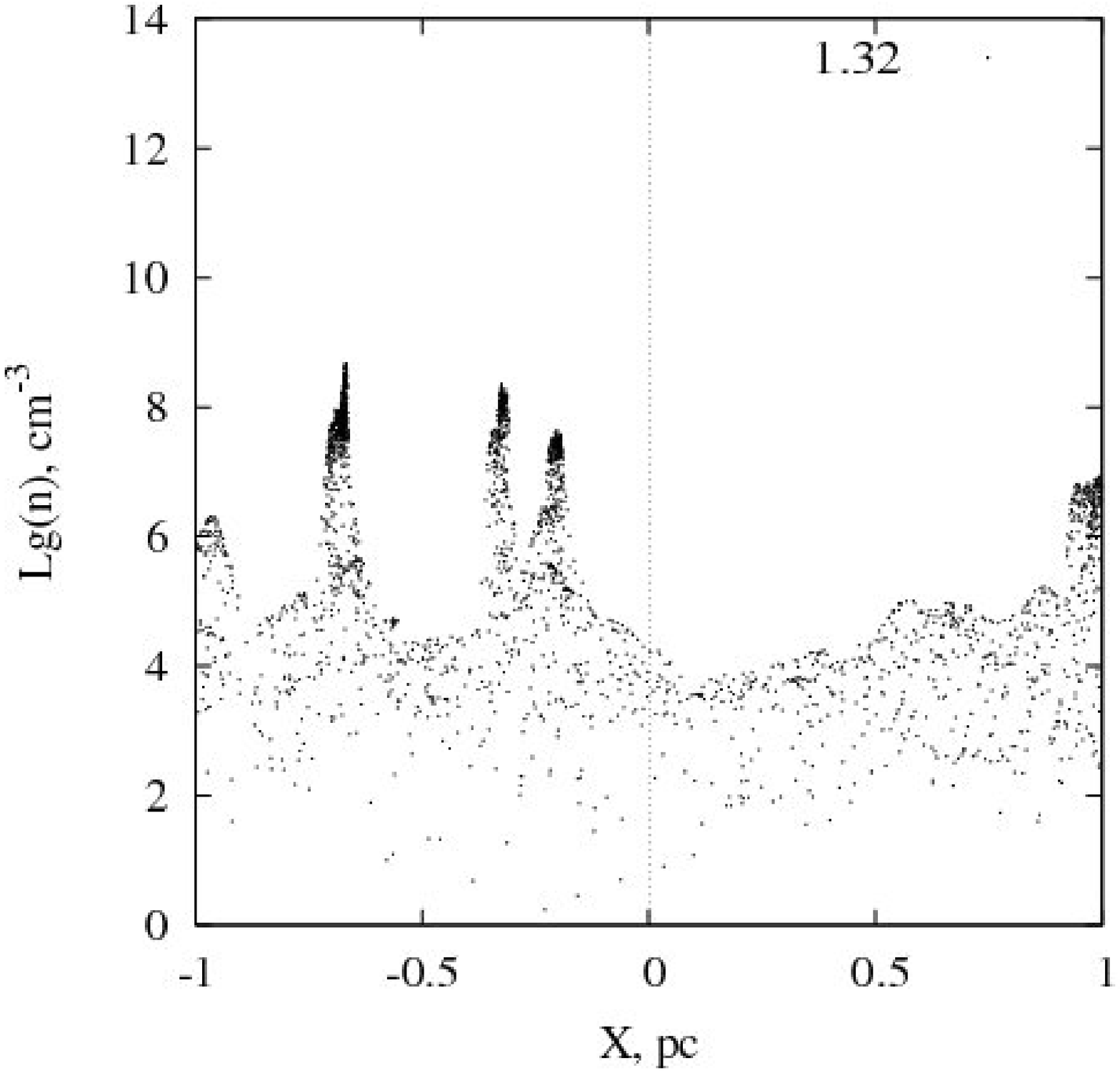}
\caption{Column-density images with $X-\lg(n)$ plots for $\beta= 0.75$.
         The top row is for time $t= 0.83 \; \textrm{Myr}$ and the bottom 
         one -- for $t= 1.32 \; \textrm{Myr}$.
         See notes on Fig. \ref{P:f-1}}
\protect\label{P:f-4}
\end{figure}

\underline{\textbf{\textsl{Collision with $\beta=0.75$}}} \hspace{1mm} 
\textsl{(see fig. \ref{P:f-4} for illustration)} \hspace{7mm}
This is largest impact distance we was considered. In difference from all 
previous models, now clumps' centers passed edgeways through forming shock front. 
Therefore, at chosen clumps velocities and such wide impact parameter the whole 
collision yield the greatest fragments amount, but with lesser degree of gas 
mixing since main shaking take place just in shock front.

As before, first intense formation of fragments appeared at $\simi 0.65 
\;\textrm{Myr}$ in the shock front and the second fragments generation had been 
forming in outer side of both lengthy arms $\simi 0.9 \;\textrm{Myr}$ later. 
But soon, almost all fragments and other gas go far from place of their "birth", 
so that the nearest to center fragments, which are found on about $6 \;\textrm{pc}$ 
from center, are relatively massive dense disc with $\mathfrak{M}= 414 \;\Msun$, 
$n \lesssim \scient{1.1}{9} \;\textrm{cm$^{-3}$}$ and smaller core in $51.5 \;\Msun$ 
with shell. The rest fragments have larger remoteness of $16 - 17 \;\textrm{pc}$ 
from center.

On both ends of extent structure are concentrated by about $1500 \Msun$ in 
fragments. There are core in $759.25 \;\Msun$ ($n \lesssim \scient{4.8}{9} 
\;\textrm{cm$^{-3}$})$ with small shell of $54.15 \;\Msun$. On the another end 
we have pair of massive fragments groups too: two dense cores with masses 
$188.25$ and $614 \;\Msun$ with shells in $591$ and $71.75 \;\Msun$ 
correspondingly. Besides there are yet a few smaller fragments here with masses 
less then $40 \Msun$ scattered around.

In simulations with worst mass resolution we have more poor fragmentation, but 
in other respects after all, there are lesser qualitative distinctions between 
simulations results for $N= 2 \mult 4000$ and $N= 2 \mult 8000$, then in three 
previous models.

\section{Discussion.} \label{S:Discussion}

At chosen clumps parameters, their masses significantly exceed Jeans' mass, 
hence they are gravitationally unstable. Theoretically each clump, taken 
separately, will collapse out during free-fall time-scale and all fragments, 
formed under contraction, in the end will have fallen onto the central object. 
But picture had been radically changed, if both clumps collide rapid enough 
one with another, so that collision time is lesser then free-fall time (in our 
models $\tau_{cross} \!:\! \tau_{f\!f} = 1\!:\!3$). Now, the evolution of forming 
fragments will depend from initial impact parameter $\beta$ or (that is same) 
from initial angular moment of whole system.

In spite of $\beta$ varying in a wide range (and independently from $N$) in 
all models began intensive gas fragmentation after nearly $\tau_{cross}$ from 
start of collision, that is in good agreement with such previous work as 
\citet{Gittins03} and \citet{Bhattal98}. The number density at this time never 
exceed opaque limit in $\simi \scient{2.6}{10} \;\textrm{cm$^{-3}$}$ (see sec. 
\ref{S:Intro}); hence the isothermal approximation is quite available, at last, 
during this period \citep[see][]{Larson69, MI2000}.

Some general information about gas fragmentation is given in tab. \ref{T:fragtab}. 
But it is necessary to note here, that owing to improvement of resolution, with 
invariable SLR and $N_{min}$ (see sec. \ref{SS:FOF}) at greater particles number 
$N$ fragments will be detect earlier and in a larger amounts, then in models 
with the worst resolution. Therefore, in the tab. \ref{T:fragtab} are given 
moments of begin intensive fragmentation in frontier shock front forming during 
collision, instead of time of first fragments detection. 

Since we have carried out comparatively low-resolution modeling we concentrate 
in this article mainly on integral characteristics of system. The main attention 
we devoted to the general properties of fragmentation -- mass spectra of 
condensation, their mass share in whole system, \etc.

There is a clear correlation between impact parameter $\beta$ and quote of 
remained fragments to its maximal quantity during evolution: $\kappa \equiv 
N_{final}/N_{max} \propto \beta$ (see tab. \ref{T:fragtab}). This hint at 
interclumps collisions may play role not only as a trigger mechanism in ISM 
fragmentation/starformation, but also as stabilizer of formed fragments system 
and set the rate of this processes.

We was not built simple, traditional differential mass function, since they 
showed most sensitiveness to chosen mass interval, quantity of fragments and so, 
then cumulative one.

The integral mass spectrum can be derived from differential one by its 
integration either in bounds $(0,m)$, that give the cumulative mass function 
$N(<m)$, i.e. the quote of fragments with masses less $m$, or in bounds 
$(m, \infty)$, give the inverse cumulative function $N(>m)$ -- 
the quote of fragments with masses greater then $m$. If differential distribution 
$\frac{dN}{dm}$ is approached by generally accepted power law $\frac{dN}{dm} = 
k m^{-\alpha}$ ($k, \alpha \!=\! \textrm{\const,} \;\alpha \!>\! 0$), then we 
have 
\[
N(<m) = \int\limits_{0}^{m} \frac{\dd N}{\dd\xi} \dd\xi = 
\frac{k}{1-\alpha} m^{1-\alpha} + c, 
\]
(where $c=$ is constant, connected with minimal fragments mass) and
\[
N(>m) = \int\limits_{m}^{\infty} \frac{\dd N}{\dd\xi} \dd\xi = 
-\frac{k}{1-\alpha} m^{1-\alpha}, \]
where we had taken into account, that $N(>m)$ is a decreasing function
\myfootnote{Of course really the integration going in finite mass interval from 
$0$ to $m_{max}$}.

\begin{figure}[Htbp]
\centerline{%
\begin{tabular}{cp{5mm}c}
\includegraphics[angle=0, width=7.5cm]{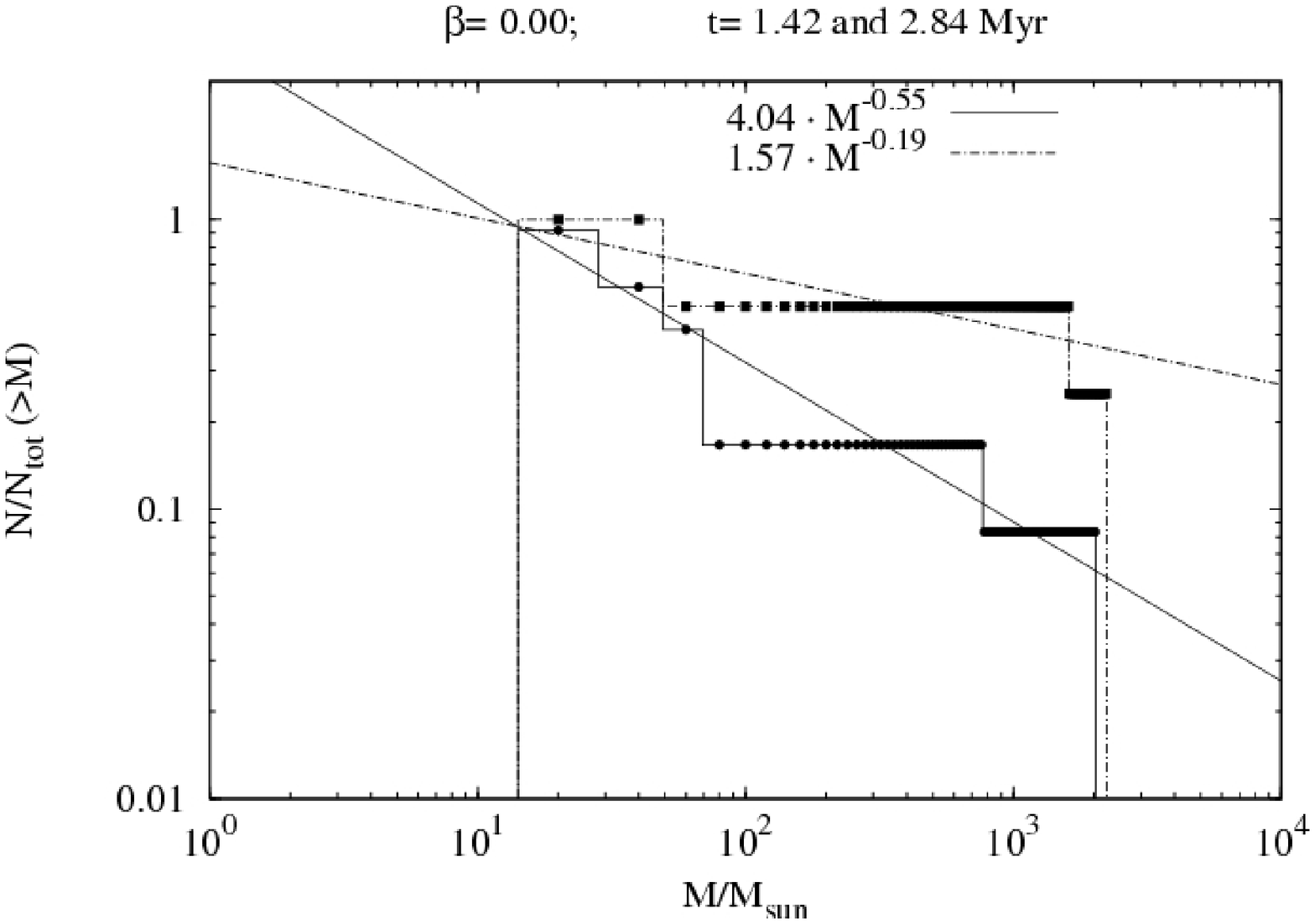} & &
\includegraphics[angle=0, width=7.5cm]{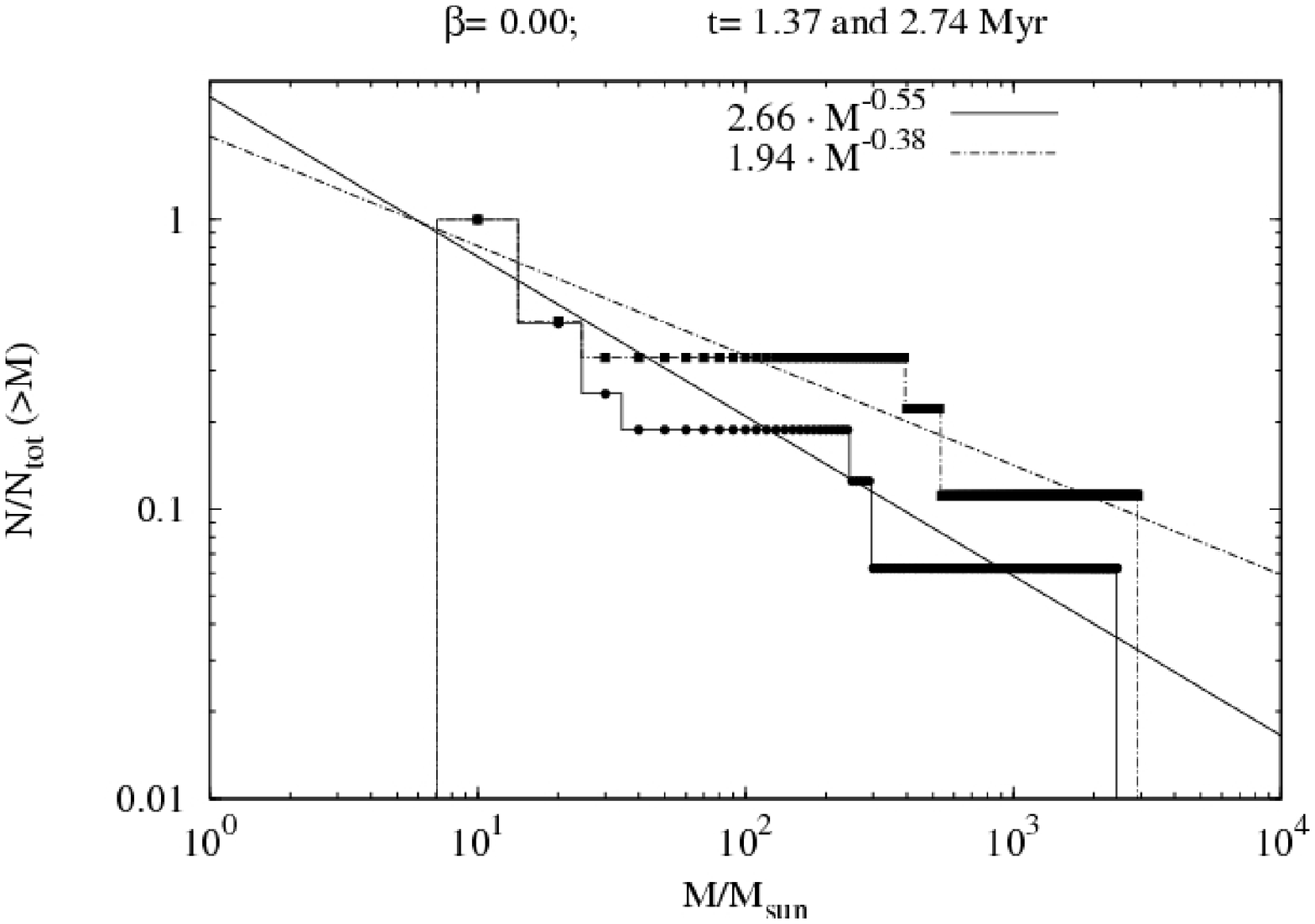} \\
\includegraphics[angle=0, width=7.5cm]{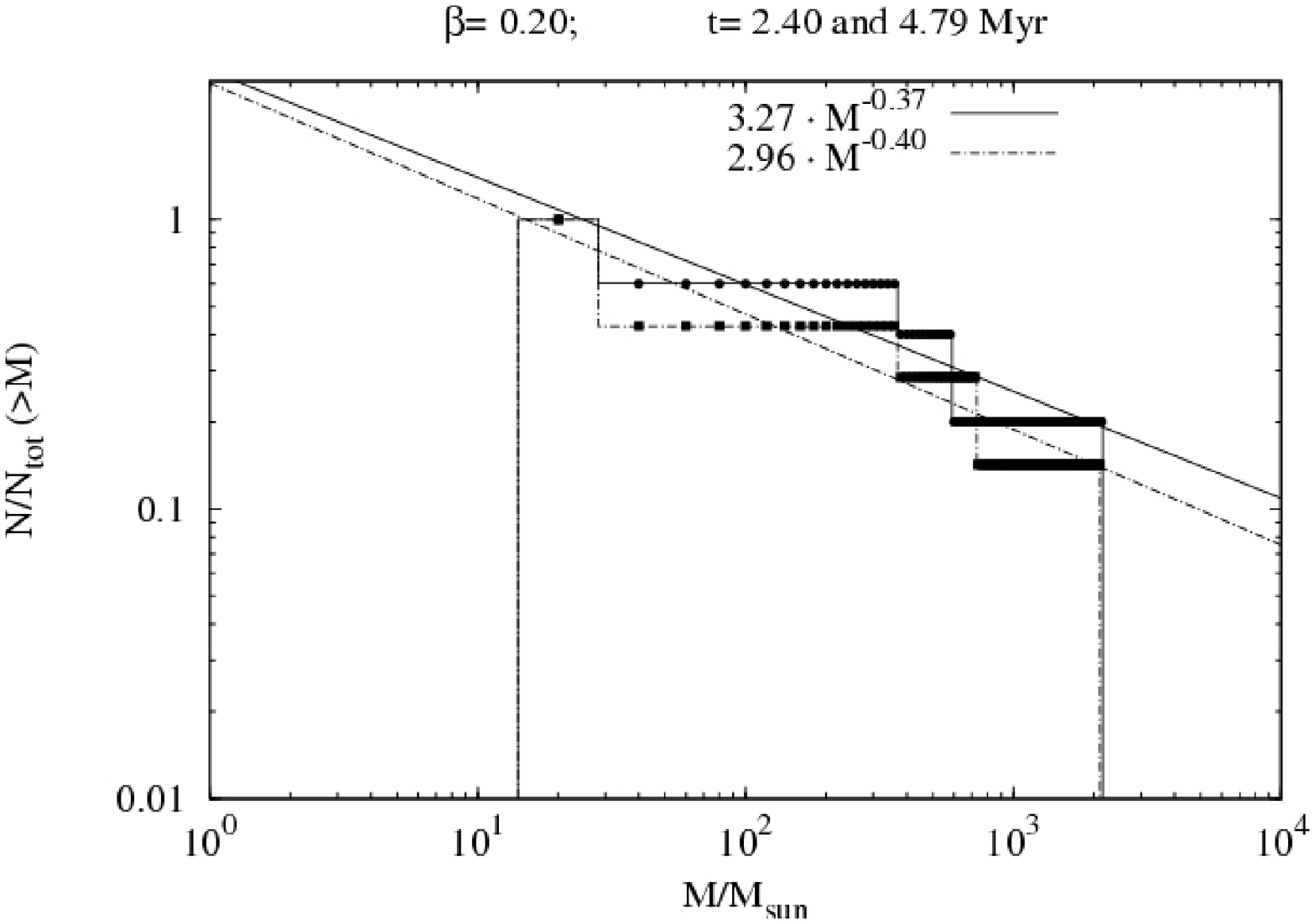} & &
\includegraphics[angle=0, width=7.5cm]{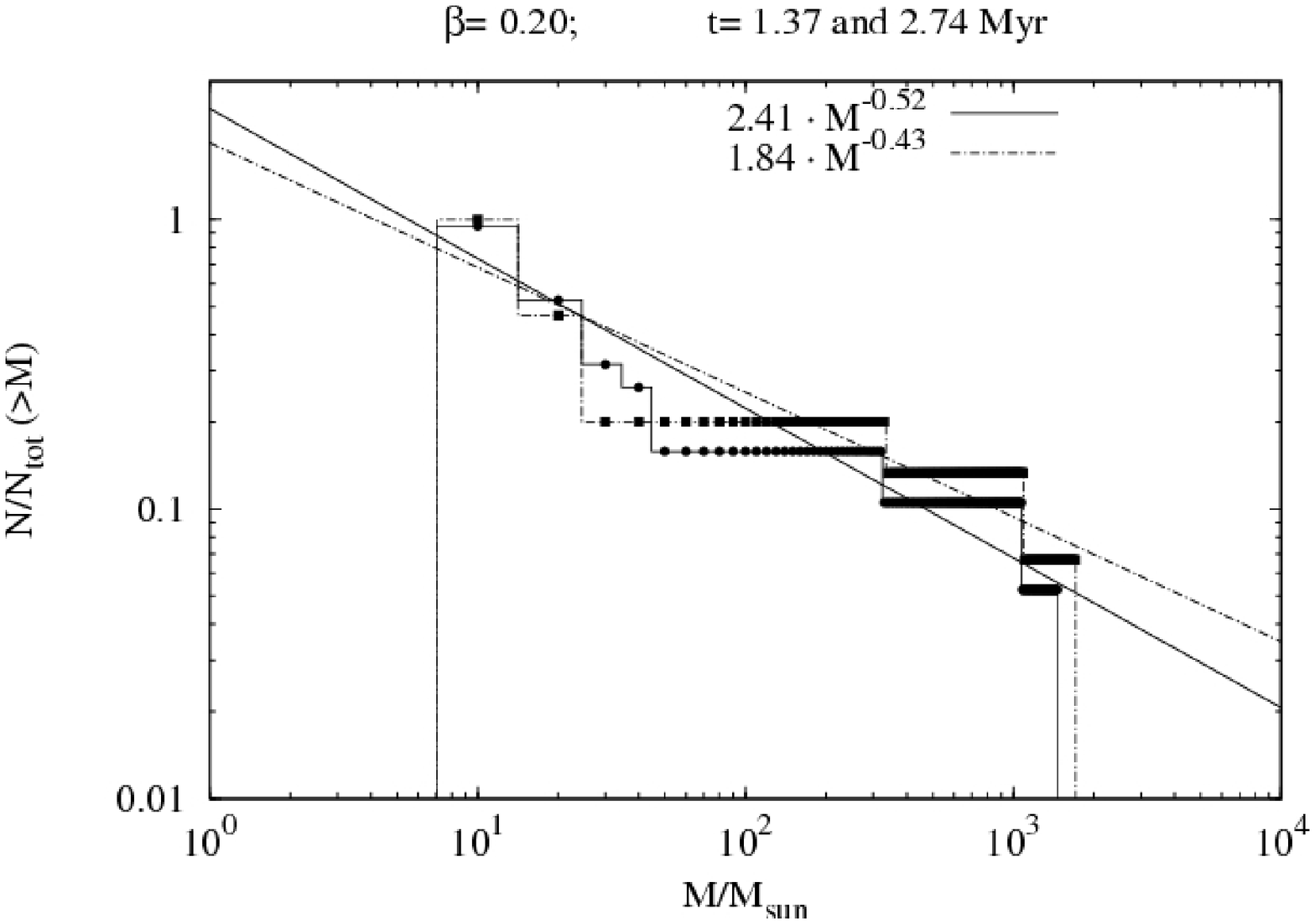} \\
\includegraphics[angle=0, width=7.5cm]{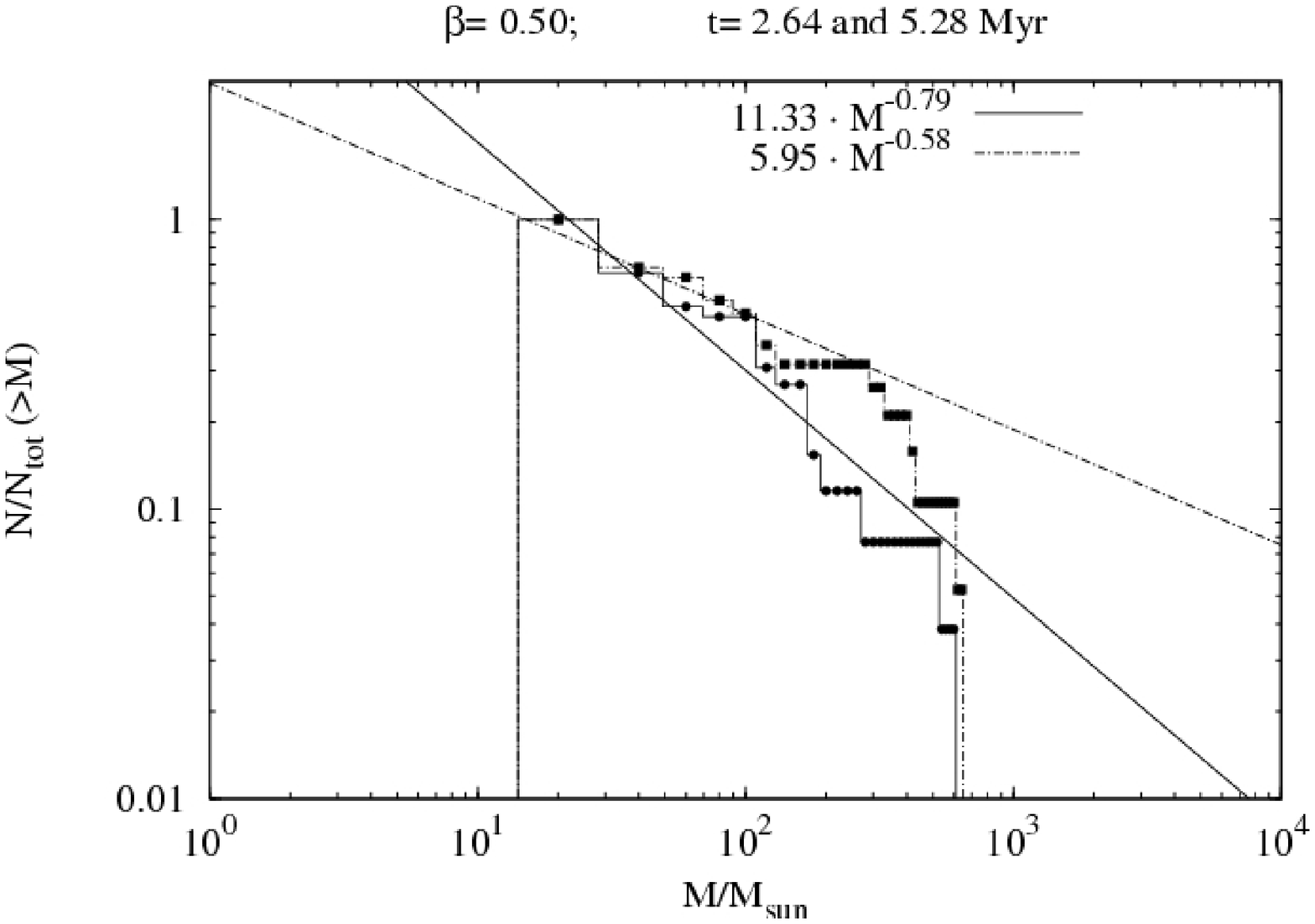} & &
\includegraphics[angle=0, width=7.5cm]{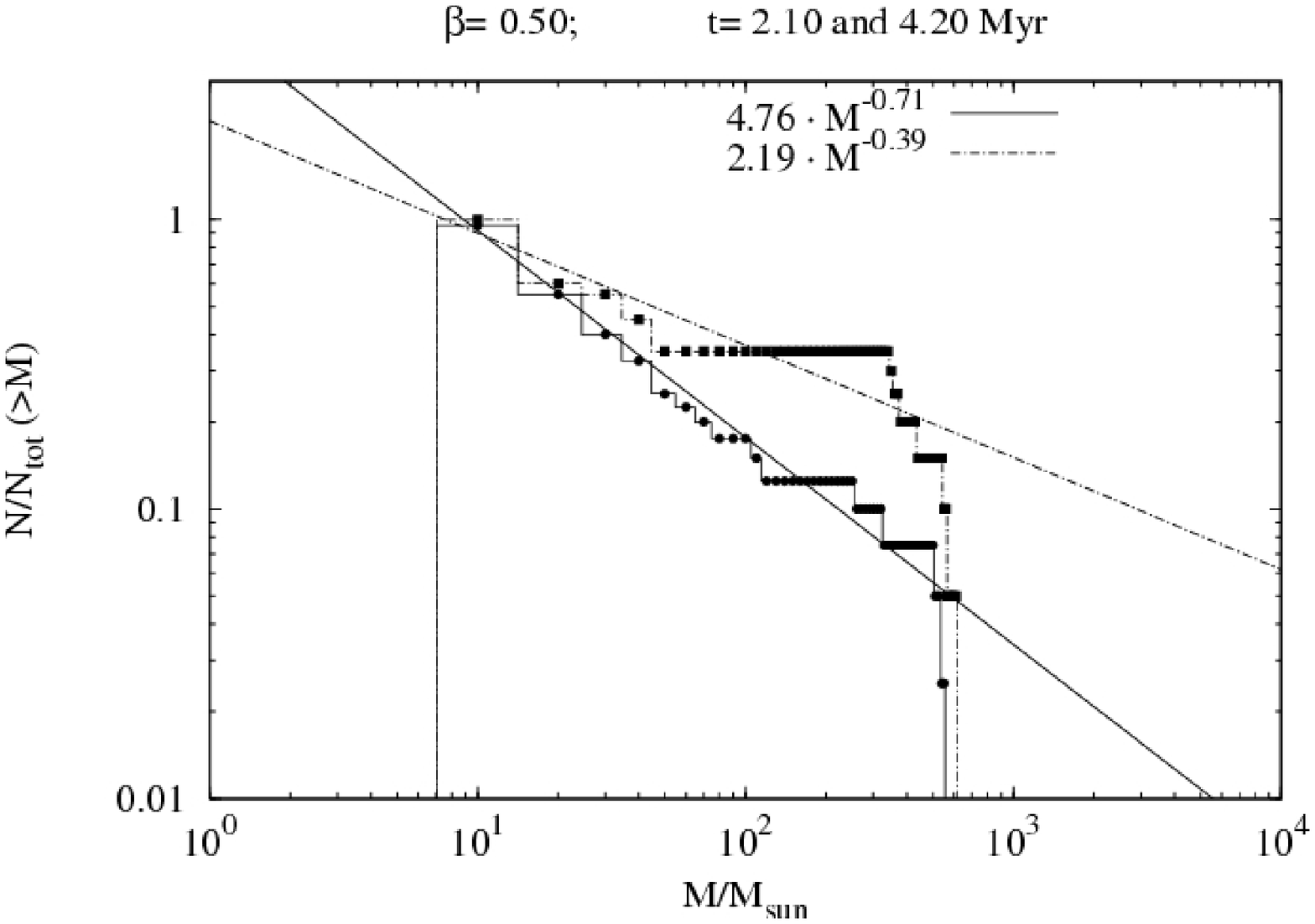} \\
\includegraphics[angle=0, width=7.5cm]{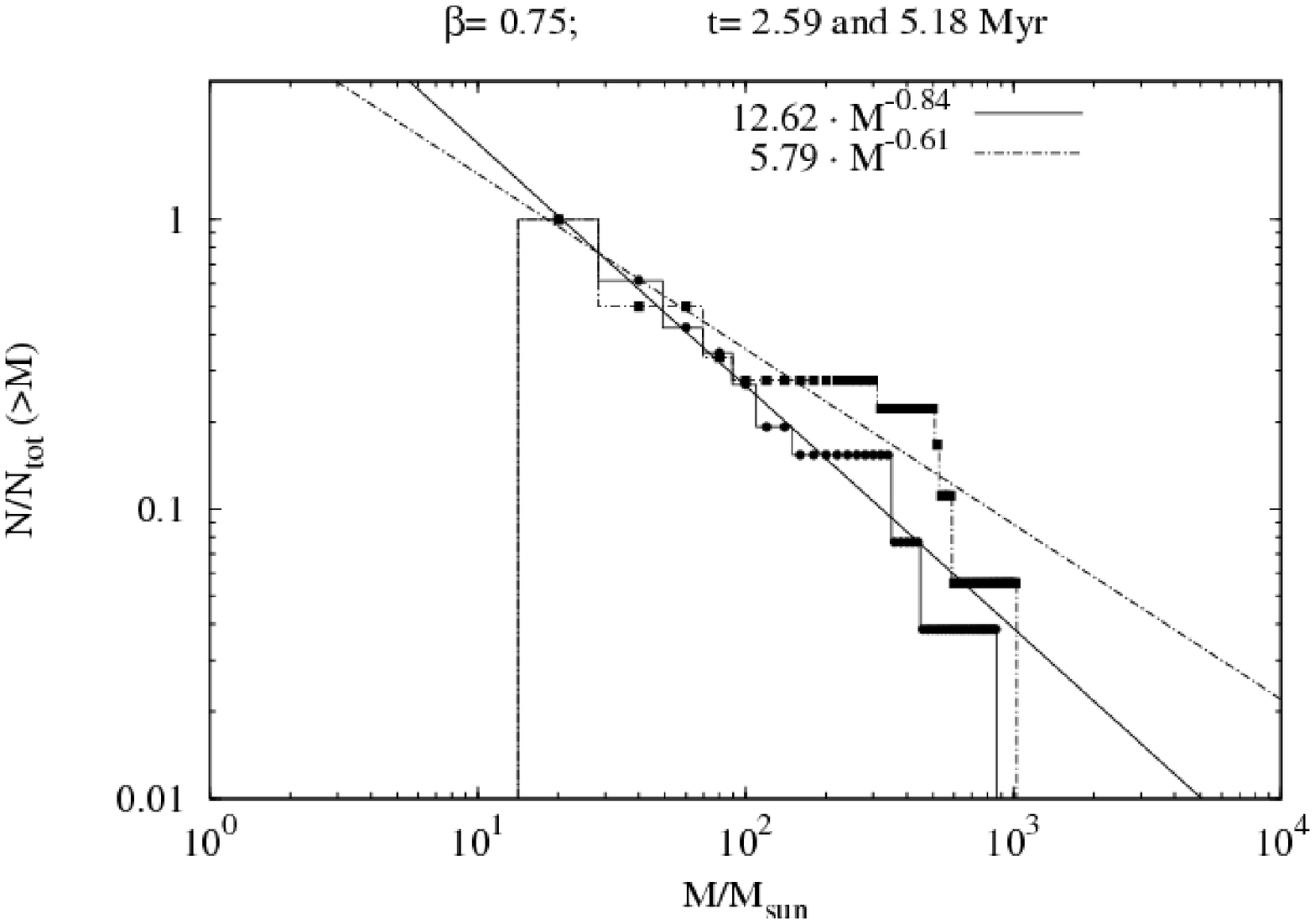} & &
\includegraphics[angle=0, width=7.5cm]{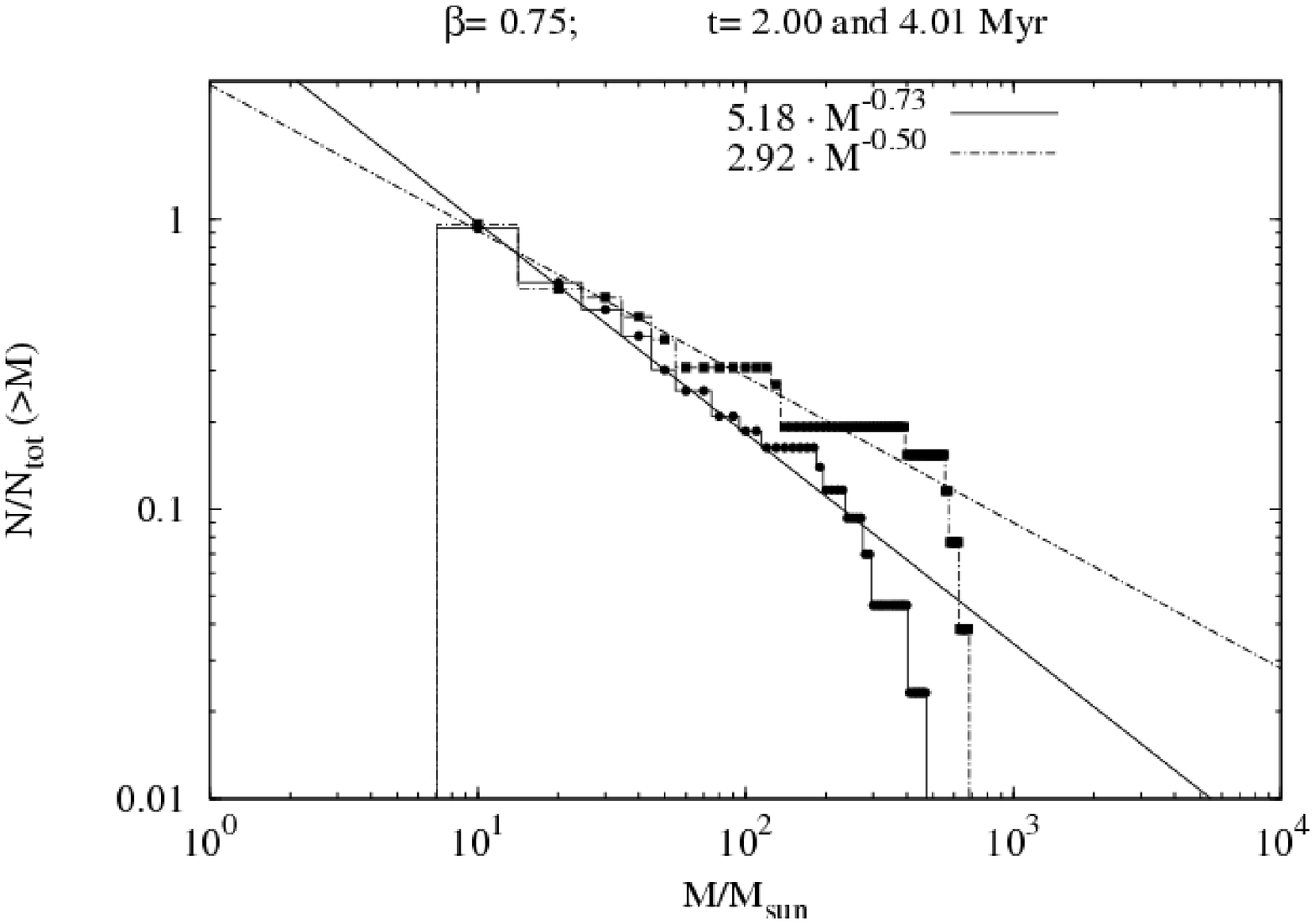} \\
\end{tabular}}
\caption{The inverse cumulative mass spectrum for models with 
         $N \!=\! 2 \times 4000$ \textit{(left)} and 
	 $N \!=\! 2 \times 8000$ \textit{(right)}.
	 Solid lines for time moment $\tau_{0.8}$ and dotted -- for 
         $2\tau_{0.8}$}
\label{P:imf}
\end{figure}

Since we have not a many data points in our spectra, we choose the inverse 
cumulative function $N(>m)\equiv1-N(<m) \propto m^{-\gamma}\; (\gamma > 0)$ for 
data fitting, as it have only two free parameters, instead three ones in 
another variant and may be linearized by transform to $log$-variables before 
fitting. The obtained mass function are given on Fig. \ref{P:imf}, and their 
slopes are summarized in tab. \ref{T:fragtab} 
~\myfootnote{However, it is necessary to note, that as we used multiscaled 
fragments searching, the massive fragments with complex internal structure 
had been split in a few lesser subfragments. Owing to this fact, the whole 
mass spectrum became flatten.}.

\begin{figure}[Htbp]
\begin{center}
\includegraphics[angle=0, width=15.0cm]{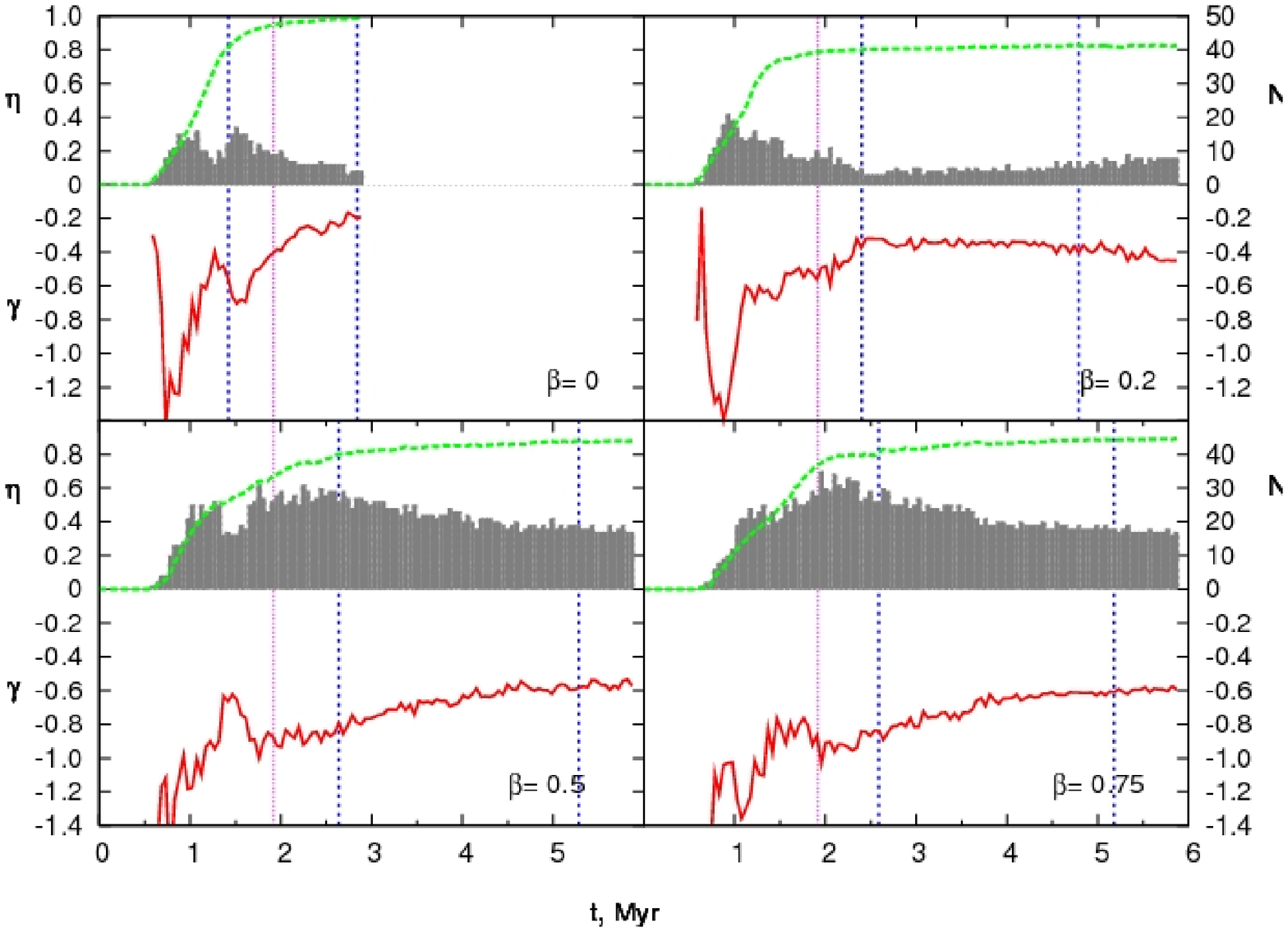} 
\includegraphics[angle=0, width=15.0cm]{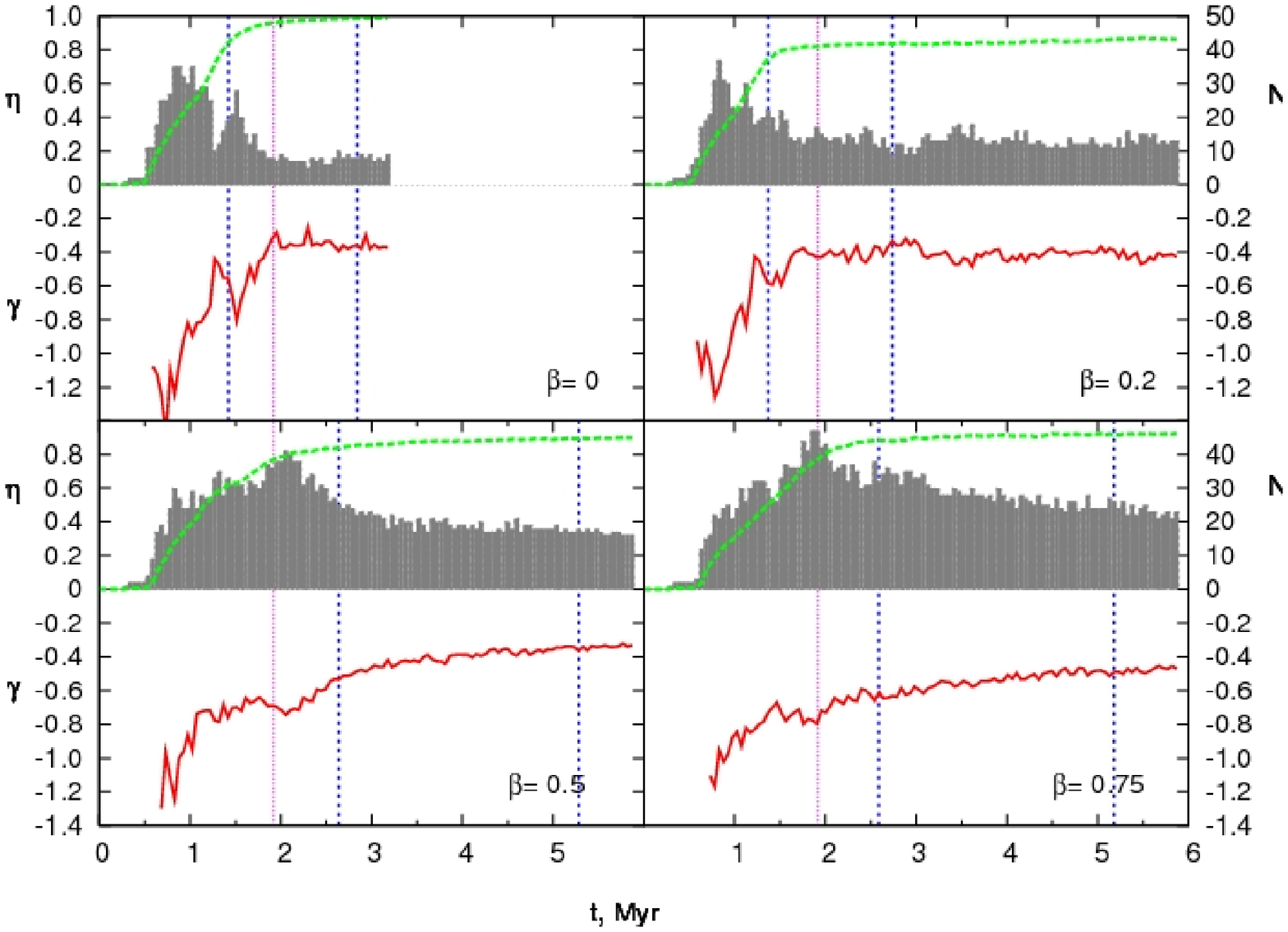} 
\end{center}
\caption{The fragments amount $N_f$ (boxes), total fragmentation $\eta$ 
         (dotted line) and slope of fragments mass spectra $\gamma$
         (solid line) \vs{} time for $N \!=\! 2 \times 4000$ \textit{(top)} 
         and $N \!=\! 2 \times 8000$ \textit{(bottom)}. The thin solid vertical 
         line and two dash-dotted lines signs the $\tau_{ff}$, $\tau_{0.8}$ 
         and $2\tau_{0.8}$ moments correspondingly.}
\label{P:fragments}
\end{figure}

The trace of fragments amount and total fragmentation index $\eta$ (that we 
defined as ratio of all fragments mass to mass of whole system: $\eta \equiv 
\mathfrak{M}_f/\mathfrak{M}$) for all evolutionary sequences are given on Fig. 
\ref{P:fragments} together with slopes $\gamma$ of fragments mass spectra. 
Above all, it is clearly seen the great similarity of curves at the same $\beta$ 
for both $N$, with the exception of differences in fragments amount on behalf 
of better resolution and limit values of $\gamma$. Also, it is seen, that 
always fragments amounts firstly strongly increase during about $0.5-0.6 
\;\textrm{Myr}$ at $\beta = 0, \; 0.2$ and $\isimi 1.5 \;\textrm{Myr}$ at 
$\beta = 0.5, \; 0.75$ up to its maximum but next begin decrease, moreover at 
small $\beta = 0.00, 0.20$ most rapid, then in two other cases. As to the 
total fragmentation $\eta$, it rapid increase together with fragments amount, 
and next its grow practically stopped, not changing to decreasing.

This fact may be explained as that intensive gas disintegration begin at 
nearly $\tau_{cross}$ and next had been broken out after a time about $1.5 - 2 
\;\textrm{Myr}$. But existing fragments continue stick one with another and 
accrete surrounding gas. Apart from this, the great fragmentation was being 
achieved at larger impact parameters, with one exception of head-on collision, 
that gave mainly total fragmentation.

Finally, we had built the dependence of cumulative mass functions slope \vs 
time for each calculated model, that is present on Fig. \ref{P:fragments} by 
solid line. 

As it seen, $\gamma$ smooth rush to certain limit, that is $-0.5 ... -0.6$ 
in models with $N= 2 \times 4000$ and close to $-0.4$ in models with 
$N=2 \times 8000$.

But, since $\gamma$ is continual vary and admissible of our isothermal 
approximation is restricted in time, we must dwell on some "typical" its values 
for represents of our results and comparisons of mass spectra slopes between 
different models. It is naturally to tie up the typical $\gamma$ with certain 
typical time moments in connection with process of fragmentation. We chose two 
such benchmark moments -- $\tau_{0.8}$, when $80\%$-fragmentation is reached, 
and $2\tau_{0.8}$ (Fig. \ref{P:imf}).

When 80\% of gas got to fragments the mass spectra slopes are nearly $0.5$ 
for first two models with $\beta= 0, \; 0.2$ and $0.7$ for two other models 
(with estimate error less then $\pm 0.03$). After yet $\tau_{0.8}$ period 
the difference between slopes for collisions with various $\beta$ had been 
rub off, all spectra became flatten and amount to $\isimi 0.4-0.6$. 

The resulting slopes of mass spectra are in good agreement with many 
observational data for molecular clouds and clumps in our Galaxy, especially 
for large collisions with $\beta= 0.5, \;\textrm{and}\; 0.75$. Thus, 
\citet{Casoli_etal84} give $\gamma= 0.59$ for $250$ clumps with masses from 
$\scient{4}{2}$ to $\scient{2}{5}$ in Perseus arm.
\citet{Simon_et_al_2001} from FCRAO and  Milky Way Galactic Ring Survey 
in ${}^{13}$CO spectral line received $\gamma= 0.80 \pm 0.10$ for central 
Galaxy region; 
\citet{Heyer_etal2002} from FCRAO Outer Galaxy Survey in lines 
${}^{12}$CO/${}^{13}$CO received $\gamma= 0.80 \pm 0.03$; 
\citet{Tatsematsu_etal1993} from CS$(1\!-\!0)$ emission for Orion A GMC gave 
slope $0.6 \pm 0.3$; 
\citet{Stutzki_Gusten1990} by analysis of high-resolution maps of M17 SW in 
C${}^{18}$O(2-1), C${}^{34}$S(2-1) and C${}^{34}$S(2-1) lines calculated 
$\gamma= 0.72 \pm 0.15$; 
\citet{Kramer_etal1998} for seven molecular clouds (also including M17 SW 
with the same slope) found mainly similar slopes $\gamma= 0.6-0.8$ in a wide 
mass range $10^{-4}-10^{4} \Msun$; 

In spite of not great number of models (in all four) they have divided onto 
two types with different characteristics -- collisions with small and large 
impact parameter ($0-0.2$ and $0.5-0.75$, correspondingly). In first group, 
owing to insufficient initial angular moment, there are single dominated by 
mass center of fragmentation with more or less developed satellites system, 
spirals \etc{} have been formed. The fragments amount, rapid decrease, after 
its maximum, but mass distribution slopes are $\isimi 0.5$ after $\tau_{0.8}$ 
and $\isimi 0.4$ after twice period.

Another group -- collisions with large $\beta$ -- lead to form essentially 
more advanced set of fragments. Now there are a few nearly equivalent centers 
of fragmentation, so each of they have its own mass dominant component and 
several satellites. The fragments amount curves are more flatten and mass 
distribution are steepen: $0.7-0.8$ and $0.5-0.6$ on moments $\tau_{0.8}$ 
and $2\tau_{0.8}$ correspondingly.

\section{Conclusions.} \label{S:Conclusion}

The main our results are follow.
\begin{enumerate}

\item After nearly $\tau_{cross}$ after start of collision in all models begin 
      intensive gas fragmentation. The fragment quantity rapid increase up to 
      its maximal value during about $0.5 \;\textrm{Myr}$ at $\beta = 0, \; 0.2$
      and $\isimi 1.5 \;\textrm{Myr}$ at $\beta = 0.5, \; 0.75$. Next, existing 
      fragments continue stick one with another, but mainly no more fragments 
      formed. Owing to this fact fragments amount decrease (more smooth at lager 
      $\beta$), but their total mass share stay almost constant, from now.

\item With the exception of head-on collision, the total gas fragmentation $\eta$ 
      increase proportional to initial impact parameter. But, at that time the
      maximal degree of fragmentation ($>98.6 \%$) had been reached in 
      head-on case.

\item The quote of remained fragments to its maximal quantity during evolution 
      $\kappa$ is direct proportional to impact parameter, what mean that 
      clouds/clumps collisions may play role not only as a trigger mechanism in 
      ISM fragmentation, but also as stabilizer of formed fragments system.
      
\item Finally, obtained mass spectra of fragments, that have typical slopes 
      $0.5-0.7$ (see above) accord with many observational data about clouds 
      and clumps in the Galaxy. The mass spectra became flatten with time, but 
      steepens with impact parameter (in time range $[\tau_{0.8};2\tau_{0.8}]$). 
      The difference between two type of collision had been rub off during 
      second $\tau_{0.8}$ period (from $\tau_{0.8}$ to $2\tau_{0.8}$) and was 
      set to $0.4-0.6$.
	
\end{enumerate}

\section{Acknowledgments.} \label{S:Acknowledgments}

Authors acknowledge the Ukrainian Fund of Fundamental Research for support
by grant {\bf 02. 07. 00132}. The numerical calculations were carried out on
SUN E10k/E15k supercomputers in \textsl{Hungarian National Supercomputing 
Center} (NIIF). In connection with that \textit{P. Berczik} acknowledge the 
NIIF for support and given machine time. Authors would like to special thank 
our colleague \textit{Tamas Maray} from NIIF for his prolonged assistance
in computers using.

\textit{P. Berczik} wish to express his thanks for the support of his work 
to the \textsl{German Science Foundation} (DFG) under SFB439 (sub-project 
B11) \textsl{''Galaxies in the Young Universe''} and (computer hardware) by 
\textsl{''Volkswagenstiftung''} (Project GRACE).

Also, authors acknowledge \textit{S. G. Kravchuk} from MAO for useful remarks 
and discussion of our results.

{}


\clearpage
\begin{table}[Hbtp] 
\centering
\scalebox{0.90}{
\begin{tabular}{|c|r|r|r|c|c|c|c|} 
\hline 
$N_{proc}$ & $N_{step}$ & $t_{total}$ & $t_{grav}$ & $t_{hydro}$ & $speed$ & $\tau_p$ & $\epsilon$ \\
\hline
 1 & 177,482 & 17,461.27 & 6,455.20 & 10,285.22 & 10.164 &  ---  &  ---  \\
\hline
 2 & 167,882 &  9,254.15 & 3,778.87 &  5,033.96 & 18.141 & 0.940 & 0.943 \\
\hline
 4 & 164,704 &  5,059.01 & 1,903.46 &  2,723.69 & 32.557 & 0.947 & 0.863 \\
\hline
 8 & 168,956 &  2,741.42 &  925.20  &  1,549.85 & 61.631 & 0.963 & 0.796 \\
\hline
16 & 167,245 &  2,170.76 &  630.79  &  1,269.27 & 77.044 & 0.934 & 0.503 \\
\hline
32 & 169,125 &  2,346.10 &  570.27  &  1,328.61 & 72.088 & 0.894 & 0.233 \\
\hline
\end{tabular} 
}
\caption{CPU consumptions for test. 
\footnotesize
$N_{proc}$-- CPUs number (SUN UltraSPARC III 1050 MHz),
$N_{step}$-- time-steps number,
$t_{total}$-- the total time consumption on whole task in seconds,
$t_{grav}$-- time consumptions on gravity calculation (including tree-building) 
             in seconds,
$t_{hydro}$ -- the same on hydrodynamics (including neighbors search), 
$speed$ -- the total calculation speed in sps (\textit{s}teps \textit{p}er \textit{s}econd),
$\tau_p$ -- the share of parallel part of code in $t_{total}$,
$\epsilon$ -- the effectiveness of CPUs loading (see text).
}
\label{T:CPUdata1} 
\end{table} 

\clearpage
\begin{table}[Htbp]
\centering
\scalebox{0.75}{
\begin{tabular}{||c||c|c|c|c||c|c|c|c|c|c||c|c||}
\hline 
 & \multicolumn{4}{c||}{begin fragmentation} & \multicolumn{6}{c||}{end of simulation} & \multicolumn{2}{c||}{$\gamma$} \\
\cline{2-13}
\Large $\beta$ & $\tau$ & $n_{max}$ & $N_f$ & $\mathfrak{M}_f$ & $\tau$ & $n_{max}$ & $N_f$ & $\mathfrak{M}_f$ & $\eta$ & $\kappa$ & $\tau_{0.8}$ & $2\tau_{0.8}$ \\
\hline 
\multicolumn{13}{||c||}{\rule{0mm}{4mm} $N= 2 \mult 4000$} \\
\hline 
0.00 & 0.59 & $\scient{4.68}{5}$ & 1 & 114.5 & 2.89 & $\scient{4.68}{12}$ &  4 & 3945.0 & 0.986 & 0.257 & 0.55 & 0.19 \\
\hline
0.20 & 0.59 & $\scient{5.20}{5}$ & 2 & 107.0 & 5.87 & $\scient{2.13}{10}$ &  8 & 3304.0 & 0.826 & 0.351 & 0.37 & 0.40 \\
\hline
0.50 & 0.59 & $\scient{4.94}{5}$ & 1 &  21.0 & 5.87 & $\scient{1.14}{11}$ & 17 & 3519.0 & 0.880 & 0.390 & 0.79 & 0.58 \\
\hline
0.75 & 0.64 & $\scient{1.04}{5}$ & 1 &  26.5 & 5.87 & $\scient{4.16}{9}$ & 17 & 3578.5 & 0.895 & 0.489 & 0.84 & 0.61 \\
\hline
\multicolumn{13}{||c||}{\rule{0mm}{4mm} $N= 2 \mult 8000$} \\
\hline
0.00 & 0.54 & $\scient{3.70}{7}$ & 11& 244.50& 2.89 & $\scient{6.68}{11}$ &  9 & 3943.75 & 0.986 & 0.235 & 0.55 & 0.38 \\
\hline
0.20 & 0.49 & $\scient{1.68}{7}$ & 3 & 41.25 & 5.87 & $\scient{1.65}{10}$ & 13 & 3451.50 & 0.863 & 0.381 & 0.52 & 0.43 \\
\hline
0.50 & 0.54 & $\scient{3.30}{7}$ & 4 & 59.75 & 5.87 & $\scient{8.98}{10}$ & 16 & 3600.50 & 0.900 & 0.548 & 0.71 & 0.39 \\
\hline
0.75 & 0.59 & $\scient{3.46}{7}$ & 3 & 48.50 & 5.87 & $\scient{9.44}{9}$  & 23 & 3698.25 & 0.925 & 0.486 & 0.73 & 0.50 \\
\hline 
\end{tabular} 
}
\caption{The parameters of fragmentation for models with both $N$.
\footnotesize
$\tau$-- time in Myr,
$n_{max}$-- the maximal number density (in cm$^{-3}$) at this moment,
$N_f$-- fragments amount in system,
$\mathfrak{M}_f$ -- mass of all fragments in $\Msun$, 
$\eta$-- system fragmentation,
$\kappa$-- the quote of remained fragments to its maximal quantity,
$\tau_{0.8}$-- the moment, when $80\%$-fragmentation is reached,
$\gamma$-- the integral mass function slope ($dN/dm \propto m^{-\gamma}$).
}
\label{T:fragtab} 
\end{table}

\clearpage

\clearpage

\clearpage

\clearpage

\clearpage

\clearpage

\clearpage

\end{document}